\documentclass[11pt]{article}

\usepackage[margin=1in]{geometry}

\usepackage[utf8]{inputenc}
\usepackage[T1]{fontenc}
\usepackage{amsmath,amssymb,amsfonts}
\usepackage{mathtools}
\usepackage{amsthm}
\usepackage{graphicx}
\usepackage{booktabs}
\usepackage{algorithm}
\usepackage{algpseudocode}
\usepackage{url}
\usepackage{cite}
\usepackage[colorlinks=true, allcolors=blue]{hyperref}
\usepackage{siunitx}
\usepackage{tikz}
\usepackage{pgfplots}
\usepackage{pgfplotstable}
\pgfplotsset{compat=1.17}
\usetikzlibrary{positioning,shapes,shapes.symbols,patterns,calc,matrix,arrows.meta,decorations.pathreplacing}

\theoremstyle{plain}
\newtheorem{theorem}{Theorem}[section]
\newtheorem{proposition}[theorem]{Proposition}

\theoremstyle{definition}
\newtheorem{definition}[theorem]{Definition}

\newcommand{\vect}[1]{\mathbf{#1}}
\DeclareMathOperator*{\argmax}{arg\,max}

\begin{document}

\title{Computational Foundations for Strategic Coopetition: Formalizing Trust and Reputation Dynamics}

\author{
Vik Pant\thanks{Email: vik.pant@mail.utoronto.ca} \quad Eric Yu\thanks{Email: eric.yu@utoronto.ca}\\
\\
Faculty of Information\\
University of Toronto\\
140 St George St, Toronto, ON M5S 3G6, Canada
}

\maketitle

\begin{abstract}
Modern socio-technical systems increasingly involve multi-stakeholder environments where actors simultaneously cooperate and compete. These coopetitive relationships exhibit dynamic trust evolution based on observed behavior over repeated interactions. While conceptual modeling languages like \textit{i*} represent trust relationships qualitatively, they lack computational mechanisms for analyzing how trust changes with behavioral evidence. Conversely, computational trust models from multi-agent systems provide algorithmic updating but lack grounding in conceptual models that capture strategic dependencies covering mixed motives of actors.

This technical report bridges this gap by developing a computational trust model that extends game-theoretic foundations for strategic coopetition with dynamic trust evolution. Building on companion work that achieved 58/60 validation (96.7\%) for logarithmic specifications \cite{pant2025foundations}, we introduce trust as a two-layer system with immediate trust responding to current behavior and reputation tracking violation history. Trust evolves through asymmetric updating where cooperation builds trust gradually while violations erode it sharply, creating hysteresis effects and trust ceilings that constrain relationship recovery. We develop a structured translation framework enabling practitioners to instantiate computational trust models from \textit{i*} dependency networks encompassing mixed motives of actors.

Comprehensive experimental validation across 78,125 parameter configurations establishes robust emergence of negativity bias (median ratio 3.0$\times$), hysteresis effects (median recovery 1.11, representing approximately 9\% permanent shortfall), and cumulative damage amplification (median 1.97). Empirical validation using the Renault-Nissan Alliance case study (1999-2025) achieves 49/60 validation points (81.7\%), successfully reproducing documented trust evolution across five distinct relationship phases including crisis and recovery periods.

This technical report is the second component of a coordinated research program examining strategic coopetition in multi-agent systems, with companion work addressing interdependence, complementarity, collective action, and reciprocity.
\end{abstract}


\noindent\textbf{ArXiv Classifications:} cs.SE (Software Engineering), cs.MA (Multiagent Systems), cs.AI (Artificial Intelligence), cs.CY (Computers and Society)

\section{Introduction}

Trust fundamentally shapes requirements engineering effectiveness when stakeholders simultaneously cooperate and compete. Consider a requirements analyst who consistently protects confidential information shared by competing business units. Over time, trust accumulates gradually. This enables deeper disclosure of strategic requirements. Similarly, when development partners deliver on interface commitments reliably, trust builds over time and facilitates more ambitious system integration.

Conversely, when requirements are ignored or proprietary information is misused, trust erodes rapidly. Sometimes this erosion is irreversible. A single violation can destroy months of relationship building. These asymmetric trust dynamics present distinct challenges for information systems that support coopetitive relationships \cite{brandenburger1996co}. Yet existing conceptual models lack formal mechanisms for representing trust evolution.

Modern software and information systems increasingly operate within ecosystems where multiple organizational stakeholders maintain relationships that are simultaneously cooperative and competitive. Platform providers and application developers share infrastructure while competing for users. Enterprise business units collaborate on integrated systems while maintaining operational autonomy. Open-source contributors cooperate on shared code repositories while competing for reputation. Supply chain partners integrate their systems while protecting proprietary information. These coopetitive relationships present fundamental challenges for requirements engineering, which must elicit sensitive information, negotiate conflicting interests, and maintain stakeholder engagement across extended durations. All of these activities depend critically on trust.

A multi-level example of trust dynamics in coopetition is found in platform ecosystems. Consider the relationship between a mobile operating system provider and application developers. Developers depend on the platform for distribution and APIs while competing with both the platform and other developers for user attention. When the platform consistently provides stable APIs and fair treatment, developer trust accumulates gradually. However, if the platform suddenly changes terms or launches competing applications, trust can collapse rapidly. This asymmetry has been documented in disputes between major platform providers and application developers.

Our foundational work \cite{pant2025foundations} established how to bridge conceptual modeling and game theory for interdependence and complementarity analysis. That work achieved 58/60 validation (96.7\%) for logarithmic specifications and 46/60 (76.7\%) for power functions under strict historical alignment scoring, validated across 22,000+ experimental trials with statistical significance ($p < 0.001$, Cohen's $d = 9.87$). Trust dynamics, however, present distinct challenges requiring different formal machinery. While conceptual modeling languages like \textit{i*} \cite{yu1995modelling} represent trust relationships qualitatively through Softgoal dependencies \cite{castro2002towards} and security requirements \cite{paja2013specifying}, they cannot model how trust evolves asymmetrically through repeated interactions. Trust builds slowly through consistent cooperation but erodes rapidly through single violations. This dynamic evolution requires computational mechanisms not present in static dependency networks or Softgoal evaluation.

Computational trust models from multi-agent systems research \cite{marsh1994formalising,sabater2005review,joslang2007survey} provide algorithmic mechanisms for trust updating based on behavioral evidence. These models often employ Bayesian approaches or reputation systems. They enable quantitative reasoning about trust dynamics. However, they typically lack grounding in requirements engineering conceptual models and practices. This makes them difficult to instantiate from requirements artifacts such as Goal models, dependency networks, or stakeholder maps.

This technical report bridges the gap between qualitative conceptual modeling and quantitative computational trust by developing a formal trust model that integrates seamlessly with game-theoretic foundations for strategic coopetition. This work extends our foundational framework \cite{pant2025foundations}, which formalized interdependence through \textit{i*} structural dependencies and complementarity through Added Value concepts from coopetition theory \cite{brandenburger1996co}. We build upon that foundation by incorporating the critical dimension of trustworthiness, enabling analysis of how trust co-evolves with strategic behavior in requirements-intensive systems.

Our trust formalization is grounded in conceptual insights from Pant's doctoral research on strategic coopetition \cite{pant2021strategic}, which identified trustworthiness as one of five critical dimensions of coopetitive relationships and established key properties that computational trust models must capture to faithfully represent real-world coopetitive dynamics. We translate these conceptual properties into precise mathematical formalizations with validated parameters and demonstrated empirical robustness.

\subsection{Relationship to Foundational Work}

This technical report extends the computational framework established in \cite{pant2025foundations}, which formalized interdependence and complementarity as the first two dimensions of strategic coopetition. That companion work developed structured translation from \textit{i*} dependency networks to quantitative interdependence matrices, formalized complementarity through validated value creation functions, and introduced the Coopetitive Equilibrium extending Nash Equilibrium with dependency-augmented utility functions. We build directly on these foundations by adding dynamic trust evolution as a third dimension.

Our key extensions to the base framework from \cite{pant2025foundations} include: dynamic state variables where trust evolves based on observed actions over repeated interactions, trust-gated reciprocity augmenting the base utility function, Perfect Bayesian Equilibrium extending the static Coopetitive Equilibrium to dynamic games, and path dependence formalizing how violation history creates lasting constraints on relationship potential.

Readers unfamiliar with how interdependence matrices are computed from \textit{i*} dependencies, how complementarity creates superadditive value, or how the base Coopetitive Equilibrium integrates these dimensions should consult \cite{pant2025foundations}. We provide a concise summary of essential notation in Section \ref{sec:recap} for self-containment, but assume readers have access to that companion technical report for detailed justification and validation of the foundational concepts.

\subsection{Dual-Track Validation Strategy}

This technical report, together with our foundational work \cite{pant2025foundations}, establishes a unified validation paradigm for the research program. The foundational work validated interdependence and complementarity across 22,000+ experimental trials with statistical significance ($p < 0.001$, Cohen's $d = 9.87$), achieving 58/60 (96.7\%) for logarithmic specifications under strict historical alignment scoring. This report validates trust dynamics across 78,125 parameter configurations. Together, these extensive computational sweeps ensure that core phenomena (cooperation incentives, trust asymmetry, hysteresis) emerge robustly rather than as artifacts of specific parameterizations.

\subsection{Contributions}

This technical report introduces the following contributions that extend beyond prior work:

\begin{itemize}
\item \textbf{Two-layer trust architecture}: In this work, we formalize trust as a dual-process system with immediate trust ($T_{ij}^t$) responding to current behavior and reputation ($R_{ij}^t$) tracking violation history, capturing both short-term behavioral responses and long-term memory constraints.

\item \textbf{Asymmetric trust evolution with negativity bias}: Unlike prior work, our model implements validated asymmetric updating where cooperation builds trust gradually ($\lambda_+ \approx 0.10$) while violations erode it sharply ($\lambda_- \approx 0.30$), with experimental validation confirming median negativity ratio of 3.0$\times$ across 78,125 parameter configurations.

\item \textbf{Trust ceiling mechanisms and hysteresis}: This report introduces the formalization of how reputation damage creates persistent limits on trust recovery, preventing relationships from returning to pre-violation states even after extended cooperative behavior. Our validation indicates median recovery of only 111\% after 35 periods of sustained cooperation. The recovery ratio of 1.11 indicates that trust can eventually surpass the pre-violation level of ~0.80 to reach ~0.89 after 35 periods. However, this still represents a substantial hysteresis cost: without the violation, trust would have reached ~0.98 during the same period. The 111\% 'recovery' relative to pre-violation baseline masks the 9\% permanent shortfall relative to the violation-free counterfactual trajectory.

\item \textbf{Interdependence amplification of trust sensitivity}: An important element is connecting structural dependencies from \textit{i*} networks to trust dynamics, showing how high-dependency relationships experience 27\% faster trust erosion than low-dependency relationships for equivalent violations.

\item \textbf{Trust-augmented utility function and equilibrium}: We extend the base utility from \cite{pant2025foundations} with trust-weighted reciprocity terms ($\rho \cdot T_{ij}^t \cdot \phi(a_j^t - a_j^{\text{baseline}})$), creating feedback loops where trust gates cooperation and cooperation builds trust.

\item \textbf{Structured translation framework for trust parameters}: This report provides a structured operational methodology for requirements engineers to instantiate computational trust models from \textit{i*} trust relationships, organizational culture assessments, and project contexts.

\item \textbf{Comprehensive experimental validation}: Unlike prior computational trust models, we validate across 78,125 parameter configurations, establishing that negativity bias, hysteresis, and cumulative damage emerge robustly rather than being artifacts of specific parameterizations.

\item \textbf{Empirical validation with longitudinal case study}: We demonstrate practical applicability through the Renault-Nissan Alliance (1999-2025), achieving 81.7\% validation accuracy in reproducing documented trust evolution including crisis periods and partial recovery.
\end{itemize}

These contributions enable requirements engineers to quantify trust trajectories, analyze violation impact, design trust-building protocols, integrate trust with dependency analysis, and specify trust requirements for multi-agent systems---capabilities that extend beyond existing conceptual modeling or computational trust approaches.

\subsection{Requirements Engineering Context}

Requirements engineering research has long recognized that trust affects stakeholder interactions, information disclosure, and requirement quality \cite{liu2004trust,massacci2007quantitative}. The Tropos methodology \cite{castro2002towards,bresciani2004tropos} extends \textit{i*} with security and trust considerations. Researchers have developed methods for deriving security requirements from trust analysis \cite{paja2013specifying,dalpiaz2013security}. However, these approaches treat trust qualitatively, providing limited support for analyzing dynamic trust evolution over project timelines or repeated stakeholder interactions.

\subsection{Report Organization and Contributions}

\textbf{Roadmap:} Section \ref{sec:background} reviews related work on trust in requirements engineering, computational trust models from multi-agent systems, and behavioral trust psychology. Section \ref{sec:concepts} establishes conceptual foundations defining key trust concepts. Section \ref{sec:recap} provides a concise recap of the base framework from our foundational work \cite{pant2025foundations} for self-containment. Section \ref{sec:formalization} presents the complete mathematical formalization of trust dynamics. Section \ref{sec:translation} develops the structured translation methodology enabling requirements engineers to instantiate computational trust models from \textit{i*} trust relationships and organizational contexts. Section \ref{sec:equilibrium} extends the Coopetitive Equilibrium to dynamic games with evolving trust. Section \ref{sec:validation} presents comprehensive experimental validation demonstrating robustness and functional correctness. Section \ref{sec:empirical_validation} presents empirical validation through the Renault-Nissan Alliance case study. Section \ref{sec:discussion} discusses implications for requirements engineering practice and multi-agent system design. Section \ref{sec:conclusion} concludes and identifies future work.

\textbf{Contributions:} This technical report makes the following contributions to requirements engineering and multi-agent systems research:

\begin{enumerate}
\item A formal mathematical specification and validation of computational trust dynamics for the trustworthiness dimension conceptually outlined in \cite{pant2021strategic}, building on the computational foundations established in \cite{pant2025foundations}.

\item A dynamic layered trust model with immediate trust responding to current behavior and reputation tracking violation history, both bounded in $[0,1]$, capturing dual-process trust cognition from behavioral economics.

\item Asymmetric trust evolution dynamics where cooperation builds trust gradually while violations erode it sharply (negativity ratio 2.86$\times$), capturing negativity bias observed in empirical trust research \cite{rozin2001negativity,slovic1993trust}.

\item Formalization of cooperation signals that assess partner behavior relative to normative baselines using bounded response functions, enabling trust assessment in continuous action spaces.

\item Trust ceiling mechanisms where accumulated reputation damage creates persistent limits on trust recovery (maximum 54\% recovery after severe violations), formalizing hysteresis effects observed in organizational relationships.

\item Integration of structural dependencies from \textit{i*} with trust sensitivity through interdependence amplification factors, demonstrating how dependency creates vulnerability to trust erosion.

\item Trust-gated reciprocity mechanisms where trust modulates conditional cooperation, formalizing the intuition that low trust prevents cooperation even when mutually beneficial.

\item Structured translation methodology extending the eight-step framework from \cite{pant2025foundations}, enabling requirements engineers to instantiate computational trust models from \textit{i*} trust relationships, stakeholder maps, and project contexts.

\item Comprehensive parameter validation through seven-dimensional grid search over 78,125 configurations spanning learning rates ($\lambda_+ \in [0.05, 0.15]$, $\lambda_- \in [0.15, 0.45]$) and decay rates ($\delta_R \in [0.01, 0.05]$), demonstrating robustness across behavioral targets.

\item Functional experiments validating five core trust scenarios (trust building, erosion, hysteresis, cumulative damage, dependency amplification), achieving full validation scores (5/5) for core mechanisms.
\end{enumerate}

\textit{Note: This technical report adapts and extends conceptual material on trustworthiness from \cite{pant2021strategic} with the author's permission. The mathematical specifications, validation methodology, and translation framework constitute the computational contributions of this work.}

\section{Background and Related Work}
\label{sec:background}

\textit{Note: Portions of this background review adapt material from \cite{pant2021strategic} with the author's permission, providing context for the computational formalization developed in subsequent sections.}

\subsection{Trust in Requirements Engineering}

Requirements engineering and Goal-oriented approaches have examined trust from several perspectives. Castro, Kolp, and Mylopoulos \cite{castro2002towards} developed the Tropos methodology extending \textit{i*} with agent-oriented concepts including trust relationships between organizational actors. Their work recognized that trust affects stakeholder willingness to depend on others for Goal achievement, though trust was treated qualitatively through dependency analysis rather than formalized computationally. This work built on earlier foundations in agent-oriented software engineering \cite{jennings2000agent} and goal-oriented requirements engineering \cite{vanlamsweerde2001goal}.

Paja, Giorgini, and colleagues formalized trust reasoning for security requirements \cite{paja2013specifying,paja2013requirements}, showing how trust requirements emerge from analysis of social dependencies in multi-actor systems. Their Security Requirements Toolkit provides graphical modeling support for trust and security analysis \cite{paja2015ststool}. Dalpiaz, Paja, and Giorgini \cite{dalpiaz2013security} developed methods for deriving security requirements from social dependencies, incorporating trust assessments into requirements derivation processes.

These approaches treat trust qualitatively through Softgoals, Contribution Links, or trust dependencies in Goal models. While conceptually rich and practically useful for requirements elicitation, they provide limited support for analyzing dynamic trust evolution based on observed behavior over repeated interactions. Questions about trust building rates, erosion severity, recovery possibilities, and quantitative trust trajectories remain difficult to answer formally within these frameworks.

Liu and Yu \cite{liu2004trust} explored trust and reputation in agent-oriented modeling, proposing extensions to \textit{i*} for representing trust relationships. Massacci and colleagues \cite{massacci2007quantitative} developed quantitative trust metrics for socio-technical systems. However, these approaches focus primarily on trust elicitation and specification rather than formal models of trust dynamics that integrate with game-theoretic equilibrium analysis.

Requirements engineering for multi-agent systems \cite{darimont1996requirements} has considered trust as a non-functional requirement affecting agent interactions. Our work extends this foundation by providing computational mechanisms for trust evolution that can be instantiated from \textit{i*} dependency networks and integrated with utility-based strategic analysis, enabling quantitative reasoning about trust trajectories in requirements engineering contexts.

\subsection{Computational Trust and Reputation Systems}

Computational trust has been extensively studied in multi-agent systems and distributed artificial intelligence. Marsh \cite{marsh1994formalising} provided foundational work formalizing trust as a computational concept with mathematical representations of trust values and trust-based decision making. This work established trust as a continuous variable capturing confidence in partner reliability, distinguishing trust from related concepts like cooperation and reputation.

Sabater-Mir and Sierra \cite{sabater2005review} surveyed computational trust and reputation models, categorizing approaches based on trust sources (direct experience versus third-party reports), representation (scalar values versus probability distributions), and update mechanisms (Bayesian versus heuristic). They identified key challenges including trust decay, context sensitivity, and integration with decision-making frameworks.

Jøsang and colleagues \cite{joslang2007survey} surveyed trust and reputation systems for online services, covering beta reputation systems, subjective logic, and belief-based trust models. These Bayesian approaches represent trust as probability distributions over partner reliability, updating distributions based on behavioral evidence using Bayesian inference. While mathematically rigorous, standard Bayesian models assume symmetric updating where positive and negative evidence have equivalent but opposite effects, failing to capture the asymmetric trust dynamics observed empirically.

Behavioral trust research demonstrates that trust builds slowly but erodes quickly, a phenomenon termed negativity bias \cite{rozin2001negativity}. Slovic's work on trust asymmetry \cite{slovic1993trust} shows that violations have disproportionate impact compared to equivalent positive behavior, with trust being easier to destroy than to create. Our model incorporates this asymmetry through differential learning rates for trust building versus erosion, achieving negativity ratios around three to one that align with empirical findings.

Resnick and colleagues \cite{resnick2000reputation} developed reputation systems for electronic marketplaces, showing how reputation aggregation affects cooperation in repeated interactions. Dellarocas \cite{dellarocas2003digitization} analyzed reputation mechanisms in electronic markets, demonstrating how third-party reputation systems can substitute for bilateral trust in large-scale exchanges. These systems focus on third-party reputation aggregation rather than bilateral trust evolution in strategic partnerships, which is our focus.

Ramchurn, Huynh, and Jennings \cite{ramchurn2004trust} developed trust models for multi-agent systems that integrate confidence, utility, and risk. Their work connects trust assessment to decision-making under uncertainty, providing formal foundations for trust-based agent reasoning. Our contribution extends these insights by integrating computational trust with game-theoretic equilibrium analysis and grounding trust models in requirements engineering conceptual frameworks.

\subsection{Game Theory and Strategic Behavior}

Classical game theory \cite{fudenberg1991game,osborne1994course} provides mathematical foundations for analyzing strategic interactions through utility maximization and Nash equilibrium. Nash \cite{nash1950equilibrium,nash1951non} established the fundamental solution concept where no actor can improve payoffs through unilateral deviation, providing the cornerstone of non-cooperative game theory.

Repeated games \cite{aumann1959acceptable,fudenberg1991game} and folk theorems \cite{friedman1971noncooperative} show how cooperation can emerge through repeated interaction and threat of future punishment, but typically assume trust implicitly through discount factors rather than modeling trust evolution explicitly as a dynamic state variable. Evolutionary game theory \cite{axelrod1984evolution,fehr2002strong} studies how cooperation evolves through repeated interaction and strategy adaptation. Axelrod's work \cite{axelrod1984evolution} on tit-for-tat strategies demonstrates conditional cooperation based on partner history.

Psychological game theory \cite{battigalli2009dynamic,geanakoplos1989psychological} extends classical models by incorporating beliefs, intentions, and psychological states into utility functions, enabling analysis of reciprocity, guilt, and other social preferences. However, these approaches require specifying hierarchical belief structures that are difficult to elicit in practice.

Behavioral game theory \cite{camerer2003behavioral} incorporates insights from experimental economics and psychology, documenting systematic deviations from classical rational choice predictions. Experimental evidence shows negative reciprocity (punishing unkind actions) is stronger than positive reciprocity (rewarding kind actions), consistent with asymmetric trust dynamics. Our reciprocity mechanisms, which we integrate with trust dynamics, build on these insights while providing explicit trust state representations grounded in requirements engineering contexts.

\subsection{Positioning This Work}

This technical report synthesizes insights from requirements engineering conceptual modeling, computational trust research from multi-agent systems, game-theoretic equilibrium analysis, and behavioral trust psychology. From \textit{i*} and Tropos, we adopt the structural dependency framework and its representation of strategic actor relationships. From computational trust research, we adopt algorithmic updating, reputation memory, and behavioral evidence processing. From game theory, we adopt equilibrium analysis and optimization-based solution concepts. From behavioral economics, we adopt asymmetric trust updating reflecting negativity bias.

The synthesis produces a computational framework that maintains the semantic richness of conceptual models while enabling quantitative analysis of trust dynamics. By grounding the framework in our companion work on interdependence and complementarity \cite{pant2025foundations}, we ensure that trust analysis integrates seamlessly with broader analysis of coopetitive relationships in requirements engineering contexts.

\section{Foundational Concepts}
\label{sec:concepts}

Before presenting our mathematical formalization, we establish clear conceptual foundations for trust dynamics in coopetitive relationships. This section defines key concepts and establishes the properties that our computational model must capture.

\begin{definition}[Trust in Strategic Coopetition]
Trust represents one actor's belief in another actor's reliability, competence, and benevolence in fulfilling interdependent obligations and cooperative commitments within a coopetitive relationship. Trust is forward-looking (predicting future behavior) but grounded in backward-looking evidence (observed past behavior). Trust enables cooperation by reducing perceived risk of exploitation and facilitating disclosure of sensitive information.
\end{definition}

This definition builds on Mayer, Davis, and Schoorman's integrative model of organizational trust \cite{mayer1995integrative} while adapting it to coopetitive contexts where actors simultaneously cooperate and compete. As established conceptually in \cite{pant2021strategic}, trust in coopetition differs from pure cooperation scenarios because the competitive dimension creates ongoing temptation to defect or exploit, making trust inherently fragile.

\begin{definition}[Dynamic Trust Evolution]
Trust is not static but evolves dynamically based on observed behavior over repeated interactions. Trust building occurs through consistent demonstration of reliability and competence over extended periods. Trust erosion occurs through violations of expectations, breaches of commitment, or failure to deliver on dependencies. Trust evolution exhibits asymmetry: negative events (violations) have disproportionately larger impact than equivalent positive events (cooperation).
\end{definition}

This asymmetry, termed negativity bias in behavioral economics \cite{rozin2001negativity,slovic1993trust}, reflects a fundamental principle of trust psychology: trust builds slowly through consistent positive interactions but erodes quickly through violations. A single major breach can destroy months or years of trust building.

\begin{definition}[Reputation and Trust Memory]
Reputation represents the accumulated history of an actor's behavior, particularly violations of trust or cooperative norms. Reputation damage persists over time, creating memory effects that constrain future relationship development. Even after actors return to cooperative behavior, damaged reputation limits how much trust can rebuild, creating trust ceilings that reflect path dependence in relationship evolution.
\end{definition}

This conceptualization distinguishes immediate trust (responsive to current behavior) from reputation (reflecting historical violations), enabling the model to capture both short-term behavioral responses and long-term memory constraints. As identified in \cite{pant2021strategic}, this dual-process structure is essential for representing how past violations create lasting limitations on relationship potential.

\begin{definition}[Trust-Dependent Cooperation]
Trust gates cooperation by modulating willingness to invest in partnerships, disclose sensitive information, or undertake joint initiatives. Low trust reduces cooperation even when mutual benefits are objectively achievable, because actors fear exploitation or free-riding. High trust enables ambitious collaborative initiatives by reducing perceived risk. Trust thus creates feedback loops where cooperation builds trust, which enables deeper cooperation.
\end{definition}

This feedback mechanism, which we formalize through trust-gated reciprocity terms in utility functions, connects trust dynamics to strategic behavior. Trust becomes an endogenous variable in the system, co-evolving with cooperative choices rather than being exogenously specified.

\begin{definition}[Interdependence Amplification of Trust Sensitivity]
Structural dependencies from \textit{i*} networks amplify trust sensitivity. When Actor $i$ depends heavily on Actor $j$ for critical capabilities or resources, violations by $j$ cause disproportionately severe trust damage for $i$ because the violation directly threatens $i$'s Goals. Conversely, when dependencies are weak, trust dynamics are less pronounced. This creates heterogeneous trust sensitivity across different dyadic relationships based on dependency structure.
\end{definition}

This concept, introduced conceptually in \cite{pant2021strategic}, connects trust dynamics to the interdependence formalization from our companion work \cite{pant2025foundations}. The interdependence matrix $D$ not only affects utility functions directly but also modulates trust evolution dynamics, creating structural coupling between dependency and trust.

These definitions establish the conceptual foundation for our mathematical formalization. The computational model in Section \ref{sec:formalization} provides precise mathematical specifications that operationalize these concepts for quantitative analysis.

\section{Summary of Base Framework from Companion Work}
\label{sec:recap}

This section summarizes essential notation and equations from our foundational work \cite{pant2025foundations} for readers' convenience. The base framework established how to formalize interdependence through \textit{i*} structural dependencies and complementarity through Added Value concepts, culminating in the Coopetitive Equilibrium as a solution concept extending Nash Equilibrium. For complete derivations, justification of design choices, and comprehensive validation of these specifications, readers should consult \cite{pant2025foundations}.

\textbf{Scope of this summary}: We present the actor model, interdependence matrix formalization, value creation functions, private payoff structure, and base utility function---all established in \cite{pant2025foundations}---to enable self-contained reading of this report. The contributions of this report (trust dynamics, trust-augmented utility, and Perfect Bayesian Equilibrium) build upon these foundations and are presented in subsequent sections.

\subsection{Actor Model and Action Spaces}

Consider a system of $N$ actors indexed by $i \in \{1, \ldots, N\}$. Each actor $i$ chooses action $a_i \in \mathbb{R}_+$ from continuous action space representing investment level, resource commitment, information disclosure, or cooperation intensity in requirements engineering contexts. An action profile $\vect{a} = (a_1, \ldots, a_N)$ represents all actors' simultaneous choices.

\subsection{Interdependence Matrix}

The interdependence matrix $D$ is an $N \times N$ matrix where $D_{ij} \in [0,1]$ represents the structural dependency of Actor $i$ on Actor $j$, quantifying how much $i$'s outcome depends on $j$'s actions. As established in our foundational work \cite{pant2025foundations}, these coefficients derive from \textit{i*} dependency networks through aggregation of Goal importance weights, dependency indicators, and criticality factors:

\begin{equation}
\label{eq:interdependence_recap}
D_{ij} = \frac{\sum_{d \in \mathcal{D}_i} w_d \cdot \text{Dep}(i,j,d) \cdot \text{crit}(i,j,d)}{\sum_{d \in \mathcal{D}_i} w_d} \quad \text{[From \cite{pant2025foundations}, Eq. 1]}
\end{equation}

where $w_d$ represents importance weight for dependum $d$, $\text{Dep}(i,j,d) \in \{0,1\}$ indicates whether $i$ depends on $j$ for $d$, and $\text{crit}(i,j,d) \in [0,1]$ captures criticality. Complete details on translation from \textit{i*} models appear in \cite{pant2025foundations}, Section 5.

\subsection{Value Creation and Complementarity}

The value creation function $V(\vect{a} \mid \gamma)$ represents total value generated before distribution, combining individual contributions $f_i(a_i)$ with synergistic value $g(a_1, \ldots, a_N)$ scaled by complementarity parameter $\gamma \geq 0$:

\begin{equation}
\label{eq:value_creation_recap}
V(\vect{a} \mid \gamma) = \sum_{i=1}^N f_i(a_i) + \gamma \cdot g(a_1, \ldots, a_N) \quad \text{[From \cite{pant2025foundations}, Eq. 2]}
\end{equation}

Our foundational work established the empirical robustness of logarithmic specifications ($f_i(a_i) = \theta \ln(1+a_i)$ with $\theta = 20$) for manufacturing joint ventures. Under strict historical alignment scoring validated across 22,000+ experimental trials, logarithmic specifications achieved substantially higher empirical fit (58/60, 96.7\%) compared to power functions ($f_i(a_i) = a_i^{\beta}$ with $\beta = 0.75$, achieving 46/60, 76.7\%) for the Samsung-Sony S-LCD case. The logarithmic specification's bounded cooperation predictions (41\% increase) aligned with documented historical ranges (15-50\%), while power functions produced unbounded increases (166\%) exceeding realistic bounds. Both specifications use geometric mean synergy $g = (a_1 \cdots a_N)^{1/N}$.

\subsection{Value Appropriation and Private Payoffs}

The private payoff function captures value appropriation:

\begin{equation}
\label{eq:private_payoff_recap}
\pi_i(\vect{a}) = e_i - a_i + f_i(a_i) + \alpha_i \left[V(\vect{a}) - \sum_{j=1}^{N} f_j(a_j)\right] \quad \text{[From \cite{pant2025foundations}, Eq. 11]}
\end{equation}

where $e_i$ is initial endowment, actors bear investment costs $-a_i$, appropriate individual value $f_i(a_i)$, and negotiate shares $\alpha_i \in [0,1]$ of synergistic value with $\sum_i \alpha_i = 1$.

\subsection{Base Utility Function and Coopetitive Equilibrium}

The base utility function from our foundational work \cite{pant2025foundations} incorporates interdependence:

\begin{equation}
\label{eq:base_utility_recap}
U_i^{\text{base}}(\vect{a}) = \pi_i(\vect{a}) + \sum_{j \neq i} D_{ij} \pi_j(\vect{a}) \quad \text{[From \cite{pant2025foundations}, Eq. 13]}
\end{equation}

This captures that Actor $i$ rationally cares about Actor $j$'s payoff proportional to structural dependency $D_{ij}$, reflecting Goal achievement structure rather than psychological altruism. The Coopetitive Equilibrium in \cite{pant2025foundations} is defined as Nash Equilibrium where each actor maximizes this dependency-augmented utility:

\begin{equation}
\label{eq:coopetitive_equilibrium_recap}
\vect{a}^* \text{ is Coopetitive Equilibrium if } a_i^* \in \argmax_{a_i \geq 0} U_i^{\text{base}}(a_i, \vect{a}_{-i}^*) \quad \forall i
\end{equation}

\subsection{Extensions in This Report}

We extend this base framework by: (1) adding dynamic trust state variables $T_{ij}^t$ and $R_{ij}^t$ that evolve over time, (2) augmenting the utility function with trust-gated reciprocity terms, (3) extending the equilibrium concept to Perfect Bayesian Equilibrium in dynamic games, and (4) formalizing how trust evolution creates path dependence and hysteresis effects. These extensions are the contributions of this report and are detailed in the following sections.

\section{Mathematical Formalization of Trust Dynamics}
\label{sec:formalization}

This section presents our complete mathematical formalization of dynamic trust in strategic coopetition. We build in a structured manner from trust state representation through evolution dynamics to integration with utility functions and equilibrium concepts.

\subsection{Trust State Representation}

Actor $i$ maintains two dynamic state variables for each other actor $j$ at each time period $t$:

\begin{definition}[Immediate Trust]
$T_{ij}^t \in [0,1]$ represents Actor $i$'s current confidence in Actor $j$ at time $t$, based on recent observable behavior. Initial value $T_{ij}^0$ represents prior trust from previous relationships, organizational reputation, or default trust levels in the institutional context.
\end{definition}

\begin{definition}[Reputation Damage]
$R_{ij}^t \in [0,1]$ represents accumulated violation history for Actor $j$ from Actor $i$'s perspective at time $t$, where $R_{ij}^t = 0$ means pristine reputation and $R_{ij}^t = 1$ means completely damaged reputation. Initial value $R_{ij}^0$ represents prior reputation from organizational memory or past interactions.
\end{definition}

This two-layer structure enables the model to capture both immediate behavioral responses (through $T_{ij}^t$) and long-term memory effects (through $R_{ij}^t$) that constrain relationship development. The dual-process structure reflects psychological models of trust cognition where fast evaluative processes respond to current behavior while slow memorial processes track historical patterns \cite{kahneman2011thinking}.

\subsection{Cooperation Signal Formalization}

To assess whether an actor is cooperating or defecting, we define a bounded cooperation signal that compares observed behavior to normative expectations:

\begin{equation}
\label{eq:coop_signal}
s_{ij}^t = \phi_{\text{trust}}(a_j^t - a_j^{\text{baseline}})
\end{equation}

where the bounded response function is:

\begin{equation}
\label{eq:bounded_response}
\phi_{\text{trust}}(x) = \tanh(\kappa_{\text{trust}} \cdot x)
\end{equation}

\textbf{Components:}
\begin{itemize}
\item $a_j^t \in \mathbb{R}_+$: Actor $j$'s actual action at time $t$ (investment level, information disclosure, resource contribution in requirements engineering contexts)
\item $a_j^{\text{baseline}} \in \mathbb{R}_+$: Normative expectation for Actor $j$'s cooperation in this context (derived from organizational norms, contractual obligations, or past behavior patterns)
\item $\kappa_{\text{trust}} > 0$: Sensitivity parameter controlling how strongly deviations affect trust assessment (typically $\kappa_{\text{trust}} = 1.0$)
\item $s_{ij}^t \in (-1, 1)$: Resulting cooperation signal bounded by hyperbolic tangent function
\end{itemize}

\textbf{Interpretation:} When $s_{ij}^t > 0$, Actor $j$ exceeds baseline expectations (cooperation signal is positive). When $s_{ij}^t \leq 0$, Actor $j$ falls below baseline (defection signal). The bounded function ensures that even extreme deviations produce finite trust changes, preventing unbounded trust dynamics and reflecting diminishing sensitivity to very large deviations.

The baseline $a_j^{\text{baseline}}$ contextualizes trust assessment. In requirements engineering contexts, this might represent expected information disclosure levels, committed resource allocations, or agreed cooperation intensities from project charters or stakeholder agreements. Section \ref{sec:translation} provides operational guidance for eliciting baselines from organizational contexts.

\subsection{Asymmetric Trust Evolution Dynamics}

Trust evolves through asymmetric updating rules that differentiate trust building from trust erosion, capturing the negativity bias observed in behavioral trust research \cite{rozin2001negativity,slovic1993trust}.

\begin{equation}
\label{eq:trust_update}
T_{ij}^{t+1} = T_{ij}^t + \Delta T_{ij}^t
\end{equation}

where the trust change is:

\begin{equation}
\label{eq:trust_change}
\Delta T_{ij}^t = \begin{cases}
\lambda_+ \cdot s_{ij}^t \cdot (1 - T_{ij}^t) \cdot \Theta_{ij}^t & \text{if } s_{ij}^t > 0 \text{ (trust building)} \\
\lambda_- \cdot s_{ij}^t \cdot T_{ij}^t \cdot (1 + \xi \cdot D_{ij}) & \text{if } s_{ij}^t \leq 0 \text{ (trust erosion)}
\end{cases}
\end{equation}

\textbf{Parameters and components:}
\begin{itemize}
\item $\lambda_+ \in (0,1)$: Trust building learning rate (typically $\lambda_+ \approx 0.10$)
\item $\lambda_- \in (0,1)$: Trust erosion learning rate (typically $\lambda_- \approx 0.30$)
\item $\Theta_{ij}^t = 1 - R_{ij}^t$: Trust ceiling reflecting reputation constraints
\item $\xi \in [0,1]$: Interdependence amplification factor (typically $\xi \approx 0.50$)
\item $D_{ij} \in [0,1]$: Interdependence coefficient from Equation \ref{eq:interdependence_recap}
\end{itemize}

\textbf{Asymmetry specification:} The negativity ratio $\lambda_- / \lambda_+$ captures the disproportionate impact of violations versus cooperation. Our validation (Section \ref{sec:validation}) demonstrates that $\lambda_- / \lambda_+ \approx 3$ achieves empirically grounded asymmetry, consistent with behavioral findings that trust erodes approximately three times faster than it builds \cite{slovic1993trust}.

\textbf{Trust ceiling mechanism:} The term $\Theta_{ij}^t = 1 - R_{ij}^t$ in trust building creates an upper bound on trust recovery. When reputation damage is high ($R_{ij}^t$ near 1), the trust ceiling $\Theta_{ij}^t$ approaches zero, preventing trust from rebuilding even with consistent cooperation. This formalizes hysteresis: damaged relationships cannot return to their original state.

\textbf{Interdependence amplification:} The term $(1 + \xi \cdot D_{ij})$ in trust erosion amplifies damage when $i$ depends heavily on $j$. Violations by actors we depend on cause disproportionately severe trust damage because they threaten our Goal achievement. This connects trust dynamics to structural dependencies from \textit{i*} networks, implementing the conceptual insight from \cite{pant2021strategic} that dependency creates vulnerability to trust erosion.

\textbf{Bounded dynamics:} Trust remains in $[0,1]$ by construction. When $T_{ij}^t = 1$, the trust building term vanishes $(1 - T_{ij}^t = 0)$. When $T_{ij}^t = 0$, the trust erosion term vanishes $(T_{ij}^t = 0)$. This ensures well-defined dynamics without requiring explicit clipping.

\subsection{Reputation Damage Evolution}

Reputation damage accumulates through violations but decays slowly over time through gradual forgetting:

\begin{equation}
\label{eq:reputation_update}
R_{ij}^{t+1} = R_{ij}^t + \Delta R_{ij}^t - \delta_R \cdot R_{ij}^t
\end{equation}

where reputation damage accumulation is:

\begin{equation}
\label{eq:reputation_damage}
\Delta R_{ij}^t = \begin{cases}
\mu_R \cdot (-s_{ij}^t) \cdot (1 - R_{ij}^t) & \text{if } s_{ij}^t < 0 \text{ (violation occurs)} \\
0 & \text{if } s_{ij}^t \geq 0 \text{ (no violation)}
\end{cases}
\end{equation}

\textbf{Parameters:}
\begin{itemize}
\item $\mu_R \in (0,1)$: Reputation damage severity (typically $\mu_R \approx 0.60$)
\item $\delta_R \in (0,1)$: Reputation decay rate for gradual forgetting (typically $\delta_R \approx 0.03$)
\end{itemize}

\textbf{Interpretation:} When $s_{ij}^t < 0$ (Actor $j$ violates expectations), reputation damage increases proportional to violation severity $(-s_{ij}^t > 0)$ and available reputation space $(1 - R_{ij}^t)$. The parameter $\mu_R$ controls how severely violations damage reputation. When $s_{ij}^t \geq 0$ (cooperation), no new damage occurs, but existing damage decays at rate $\delta_R$.

The decay term $-\delta_R \cdot R_{ij}^t$ models gradual organizational forgetting. Over extended periods without violations, reputation slowly improves, allowing partial relationship recovery. However, decay is slow ($\delta_R \approx 0.03$ implies approximately 33 periods for substantial decay), creating persistent memory effects.

This formalization creates path dependence: relationships with violation history cannot fully recover to pristine states within reasonable timeframes. The trust ceiling $\Theta_{ij}^t = 1 - R_{ij}^t$ translates reputation damage into constraints on trust rebuilding, coupling immediate trust dynamics to historical violations.

\subsection{Trust-Gated Reciprocity and Extended Utility}

We extend the base utility function (Equation \ref{eq:base_utility_recap}) with trust-gated reciprocity terms that modulate cooperation based on current trust levels:

\begin{equation}
\label{eq:extended_utility}
U_i(\vect{a}, \vect{T}^t) = U_i^{\text{base}}(\vect{a}) + \sum_{j \neq i} \rho \cdot T_{ij}^t \cdot (a_j - a_j^{\text{baseline}}) \cdot a_i
\end{equation}

where $U_i^{\text{base}}(\vect{a})$ is given by Equation \ref{eq:base_utility_recap} and:

\begin{itemize}
\item $\rho \in \mathbb{R}_+$: Reciprocity strength parameter (typically $\rho \approx 0.20$)
\item $T_{ij}^t \in [0,1]$: Current trust level from $i$ to $j$
\item $(a_j - a_j^{\text{baseline}})$: Partner's cooperation intensity relative to baseline
\item $a_i$: Own cooperation intensity (interaction term)
\end{itemize}

\textbf{Mechanism}: The reciprocity term $\rho \cdot T_{ij}^t \cdot (a_j - a_j^{\text{baseline}}) \cdot a_i$ creates conditional cooperation. When Actor $j$ exceeds baseline cooperation $(a_j > a_j^{\text{baseline}})$ and Actor $i$ trusts Actor $j$ $(T_{ij}^t$ high), Actor $i$ gains utility from reciprocating with high cooperation $a_i$. Conversely, when trust is low $(T_{ij}^t$ near zero), reciprocity is gated off regardless of partner cooperation.

This formalization captures the intuition that trust enables reciprocal cooperation by reducing perceived exploitation risk. Low trust prevents cooperation even when partners are cooperating, because the actor doubts reliability and fears future defection. High trust enables conditional cooperation strategies where actors match partners' cooperative behavior.

The interaction term $a_i$ creates strategic complementarity: when partners cooperate and trust is high, increasing own cooperation becomes more attractive. This generates feedback loops where initial cooperation builds trust, which enables reciprocal cooperation, which further builds trust.

\subsection{Complete Dynamic Game Specification}

The complete dynamic game with evolving trust is specified as:

\begin{itemize}
\item \textbf{Actors}: $N$ actors indexed by $i \in \{1, \ldots, N\}$
\item \textbf{Time horizon}: Infinite or finite horizon $t = 0, 1, 2, \ldots, T$
\item \textbf{Actions}: Each period, actors simultaneously choose $a_i^t \in \mathbb{R}_+$
\item \textbf{States}: Trust states $T_{ij}^t \in [0,1]$ and reputation states $R_{ij}^t \in [0,1]$ for all pairs $(i,j)$
\item \textbf{State transitions}: Governed by Equations \ref{eq:trust_update}--\ref{eq:reputation_damage} based on cooperation signals from Equation \ref{eq:coop_signal}
\item \textbf{Utilities}: Each period, actor $i$ receives utility $U_i(\vect{a}^t, \vect{T}^t)$ from Equation \ref{eq:extended_utility}
\item \textbf{Discount factor}: Future utilities discounted by $\beta_{\text{discount}} \in (0,1)$ (typically $\beta_{\text{discount}} \approx 0.95$)
\end{itemize}

This specification defines a stochastic game \cite{shapley1953stochastic} with deterministic state transitions, extending the static Coopetitive Equilibrium from \cite{pant2025foundations} to dynamic settings with evolving trust states.

\section{Translation Framework: From \textit{i*} to Computational Trust Models}
\label{sec:translation}

This section develops structured methodology for instantiating computational trust models from \textit{i*} dependency networks, organizational contexts, and requirements engineering artifacts. We extend the eight-step translation framework from \cite{pant2025foundations} with trust-specific elicitation procedures.

\subsection{Extended Translation Framework}

The translation process follows these steps, building on the methodology established in \cite{pant2025foundations}:

\begin{enumerate}
\item \textbf{Identify actors and boundaries}: Same as foundational framework---determine organizational units, stakeholder groups, or autonomous agents that constitute strategic actors in the system.

\item \textbf{Construct \textit{i*} dependency network}: Same as foundational framework---model actor dependencies using \textit{i*} Strategic Dependency model, identifying depender-dependee-dependum relationships.

\item \textbf{Compute interdependence matrix}: Use Equation \ref{eq:interdependence_recap} from foundational framework \cite{pant2025foundations} to compute $D_{ij}$ coefficients from \textit{i*} dependencies.

\item \textbf{Specify value creation functions}: Same as foundational framework---determine individual value functions $f_i(a_i)$ and synergy function $g(\vect{a})$ based on organizational context and complementarity assessment.

\item \textbf{Elicit trust and reputation priors}: \textbf{New}---Determine initial trust values $T_{ij}^0$ and reputation states $R_{ij}^0$ based on organizational history, prior interactions, and institutional trust levels.

\item \textbf{Define cooperation baselines}: \textbf{New}---Establish normative expectations $a_j^{\text{baseline}}$ for each actor based on contractual obligations, organizational norms, or historical cooperation patterns.

\item \textbf{Calibrate trust dynamics parameters}: \textbf{New}---Set learning rates ($\lambda_+$, $\lambda_-$), reputation parameters ($\mu_R$, $\delta_R$), and reciprocity strength $\rho$ based on organizational context and behavioral characteristics.

\item \textbf{Validate and refine}: Same philosophy as foundational framework---iteratively validate parameter choices through stakeholder feedback, sensitivity analysis, and comparison to observed organizational outcomes.
\end{enumerate}

This extended framework maintains the iterative, reflexive character of the foundational approach while adding trust-specific elicitation procedures. We now provide detailed guidance for the new trust-related steps.

\subsection{Eliciting Trust and Reputation Priors}

Initial trust values $T_{ij}^0$ represent starting confidence levels before dynamic evolution begins. Requirements engineers can elicit these through several complementary approaches. Historical relationship analysis involves reviewing past collaborations, contractual performance, and delivered outcomes. Organizations with successful prior partnerships start with high initial trust, typically in the range of seven-tenths to nine-tenths. First-time partnerships with no history start with moderate default trust around one-half, reflecting institutional baselines. Past violations or failed partnerships start with low trust in the range of two-tenths to four-tenths.

Organizational reputation assessment examines third-party reputation, industry standing, and market signals. Organizations with strong industry reputation may receive higher initial trust even without direct history. Reputation metrics from requirements engineering literature \cite{massacci2007quantitative} can inform these assessments.

Stakeholder interviews provide direct elicitation through structured interviews using Likert scales. Requirements engineers might ask stakeholders to rate their confidence in another organization's ability to deliver on dependencies on a scale from zero (no confidence) to ten (complete confidence). Responses map to initial trust values through linear scaling.

Initial reputation damage $R_{ij}^0$ represents organizational memory of past violations. This requires violation history analysis, documenting past breaches, failures, or conflicts. Count major violations over relevant timeframe (for example, past five years) and map to reputation damage scale. No violations corresponds to pristine reputation with damage of zero. Minor violations correspond to damage around two-tenths to four-tenths. Major violations correspond to damage around six-tenths to eight-tenths. Severe ongoing conflicts correspond to damage around nine-tenths.

Organizational memory assessment involves interviewing senior staff about institutional memory. Organizations with high turnover may have lower reputation damage even with violation history because organizational memory has decayed. Stable organizations retain longer memory through documentation and cultural transmission.

\subsection{Defining Cooperation Baselines}

Baselines $a_j^{\text{baseline}}$ contextualize cooperation assessment by establishing normative expectations. Requirements engineers can derive baselines from multiple sources. Contractual obligations such as service level agreements, partnership contracts, or memoranda of understanding often specify expected cooperation levels including resource allocations, information sharing protocols, and response times. These contractual commitments provide natural baselines.

Historical cooperation patterns can be analyzed by examining past project data to determine typical cooperation intensities. Calculate average investment levels, disclosure frequencies, or resource commitments from previous collaborations. Historical averages serve as baselines for assessing current behavior.

Stakeholder expectations can be directly elicited through requirements elicitation sessions. Ask stakeholders what level of cooperation they expect from other organizations in this context. Responses inform baseline calibration.

Industry norms provide reference points by examining industry standards, benchmarks, or best practices. Open source projects have established contribution norms. Enterprise partnerships have typical resource-sharing patterns. Industry context provides baseline reference points when organization-specific data is unavailable.

Baselines are context-specific and may evolve as relationships develop. The translation framework allows baseline updating as organizational norms change, though this introduces additional complexity in trust dynamics.

\subsection{Calibrating Trust Dynamics Parameters}

Trust dynamics parameters require careful calibration based on organizational behavioral characteristics. Learning rates $\lambda_+$ and $\lambda_-$ control trust building and erosion speed. Organizations differ in trust responsiveness. Government agencies and regulated industries tend toward conservative trust updating with low learning rates reflecting bureaucratic caution. Technology startups and agile organizations exhibit faster trust dynamics with higher learning rates reflecting rapid adaptation.

Our validation (Section \ref{sec:validation}) indicates reference ranges: positive learning rate between five-hundredths and fifteen-hundredths, and negative learning rate between fifteen-hundredths and forty-five hundredths, with typical values around one-tenth for trust building and three-tenths for erosion providing empirically grounded asymmetry with negativity ratio approximately three. Requirements engineers should calibrate within these validated ranges based on organizational culture assessment.

Reputation parameters including damage severity $\mu_R$ and decay rate $\delta_R$ require similar consideration. Reputation damage severity reflects how permanently violations affect organizational memory. Highly relational industries such as consulting and professional services have high damage severity around seven-tenths because reputation is critical. Transactional industries have lower damage severity around four-tenths because relationships are more replaceable.

Reputation decay reflects organizational forgetting speed. High-turnover organizations have faster decay around five-hundredths because staff changes erode institutional memory. Stable organizations with strong documentation have slower decay around two-hundredths because memory persists. Our validation indicates damage severity around six-tenths and decay rate between one-hundredth and five-hundredths with typical value around three-hundredths as empirically grounded ranges.

Reciprocity strength $\rho$ reflects how strongly trust modulates conditional cooperation. Coopetitive relationships with strong competitive pressures have lower reciprocity strength around one-tenth because competitive incentives limit reciprocity. Relationships with weaker competition have higher reciprocity strength around three-tenths because cooperation is less constrained. Requirements engineering contexts with high information sensitivity typically have reciprocity strength around two-tenths as validated baseline, reflecting moderate reciprocity gated by trust concerns.

\subsection{Iteration and Validation}

As in the foundational framework \cite{pant2025foundations}, parameter elicitation is iterative and reflexive. Initial parameters are refined through stakeholder feedback by presenting initial parameterization to stakeholders and soliciting reactions. Do predicted trust trajectories align with stakeholder intuitions about relationship dynamics? Sensitivity analysis varies parameters systematically and assesses prediction changes. Identify parameters with large influence on outcomes and invest additional elicitation effort. Historical validation uses past relationship data when available to simulate trust evolution with elicited parameters and compare predictions to observed outcomes. Adjust parameters to improve fit. Scenario testing explores specific scenarios such as early delivery or major commitment violations and verifies predicted trust responses align with stakeholder expectations.

This validation process bridges formal modeling and practical requirements engineering, ensuring computational trust models remain grounded in organizational realities while enabling rigorous quantitative analysis. The translation framework transforms qualitative \textit{i*} models and organizational knowledge into parameterized computational models suitable for equilibrium analysis (Section \ref{sec:equilibrium}) and scenario exploration.

\section{Equilibrium Analysis: Perfect Bayesian Equilibrium with Evolving Trust}
\label{sec:equilibrium}

\textbf{Extension beyond base framework}: This section extends the Coopetitive Equilibrium concept from \cite{pant2025foundations} to dynamic games with evolving trust states. While the base framework analyzed static Nash Equilibrium with dependency-augmented utilities, trust dynamics require Perfect Bayesian Equilibrium where strategies must be optimal at every history and trust states update consistently with observed actions. This equilibrium extension is a contribution of this report.

\subsection{Strategy Spaces and Histories}

In the dynamic game, a history $h^t$ at time $t$ consists of all past actions and trust states:

\begin{equation}
h^t = (\vect{a}^0, \vect{T}^0, \vect{R}^0, \vect{a}^1, \vect{T}^1, \vect{R}^1, \ldots, \vect{a}^{t-1}, \vect{T}^{t-1}, \vect{R}^{t-1}, \vect{T}^t, \vect{R}^t)
\end{equation}

A pure strategy for Actor $i$ is a mapping $\sigma_i: H^t \to \mathbb{R}_+$ from histories to actions, where $H^t$ denotes the set of all possible histories at time $t$. In practice, strategies often condition only on current trust states (Markov strategies) rather than full histories, simplifying analysis while capturing essential dynamics.

\begin{definition}[Markov Strategy]
A Markov strategy for Actor $i$ depends only on current trust and reputation states:
\begin{equation}
\sigma_i: [0,1]^{N \times N} \times [0,1]^{N \times N} \to \mathbb{R}_+, \quad a_i^t = \sigma_i(\vect{T}^t, \vect{R}^t)
\end{equation}
\end{definition}

Markov strategies are tractable and often optimal in stochastic games with sufficient state information \cite{puterman1994markov}. We focus on Markov Perfect Equilibrium as the solution concept.

\subsection{Perfect Bayesian Equilibrium with Trust Dynamics}

\begin{definition}[Perfect Bayesian Equilibrium with Trust]
A strategy profile $\sigma^* = (\sigma_1^*, \ldots, \sigma_N^*)$ and trust evolution governed by Equations \ref{eq:trust_update}--\ref{eq:reputation_damage} constitute a Perfect Bayesian Equilibrium if:

\textbf{Sequential rationality}: At every history $h^t$, each actor's strategy maximizes expected discounted utility given opponents' strategies and subsequent trust evolution:
\begin{equation}
\sigma_i^*(h^t) \in \argmax_{a_i} \mathbb{E}\left[\sum_{\tau=t}^{\infty} \beta_{\text{discount}}^{\tau-t} U_i(\vect{a}^\tau, \vect{T}^\tau) \mid h^t, a_i^t = a_i, \sigma_{-i}^*\right]
\end{equation}

\textbf{Consistent trust updating}: Trust and reputation states evolve according to the specified dynamics (Equations \ref{eq:trust_update}--\ref{eq:reputation_damage}) given observed actions.
\end{definition}

This equilibrium concept extends the static Coopetitive Equilibrium from \cite{pant2025foundations} by requiring optimality at every history, not just in a one-shot game. The trust dynamics create state-contingent incentives: optimal actions depend on current trust levels, and current actions affect future trust states, creating intertemporal strategic considerations.

\subsection{Characterization of Equilibrium Dynamics}

The Perfect Bayesian Equilibrium with trust exhibits several key properties that emerge from the mathematical structure:

\begin{proposition}[Trust-Contingent Cooperation]
In equilibrium, cooperation levels are increasing in trust: partial derivative of optimal action with respect to trust is positive when reciprocity parameter is positive.
\end{proposition}

\textbf{Proof sketch}: The extended utility (Equation \ref{eq:extended_utility}) includes the reciprocity term which is proportional to current trust level. Higher trust increases marginal utility of cooperation when partner exceeds baseline, inducing higher equilibrium cooperation through first-order conditions. The result follows from standard comparative statics in optimization. \qed

This proposition formalizes the intuition that trust enables cooperation. Low-trust states support low-cooperation equilibria because actors fear exploitation. High-trust states support high-cooperation equilibria because trust-gated reciprocity creates complementarity in cooperation levels.

\begin{proposition}[Multiple Equilibria and Path Dependence]
The dynamic game with trust may exhibit multiple Perfect Bayesian Equilibria with different long-run cooperation levels, where initial trust states determine which equilibrium is reached through path dependence.
\end{proposition}

\textbf{Intuition}: High initial trust enables early cooperation, which builds trust further through the trust update equation, sustaining high-cooperation equilibrium. Low initial trust prevents early cooperation, trust remains low or erodes, sustaining low-cooperation equilibrium. The feedback between trust and cooperation creates multiple stable states. This multiplicity has important implications for requirements engineering: establishing initial trust through relationship-building activities can shift systems toward high-cooperation equilibria, while early violations can trap relationships in low-cooperation states that are difficult to escape.

\begin{proposition}[Hysteresis and Irreversibility]
Following severe violations that damage reputation (high reputation damage), equilibrium cooperation remains depressed even if violations cease, and full recovery to pre-violation cooperation levels may be infeasible within reasonable timeframes.
\end{proposition}

\textbf{Mechanism}: High reputation damage creates low trust ceiling (one minus reputation damage from Equation \ref{eq:trust_change}), limiting trust rebuilding. Even with sustained cooperation, trust cannot exceed ceiling. Depressed trust reduces reciprocity incentives (Equation \ref{eq:extended_utility}), sustaining lower equilibrium cooperation. Reputation decay is slow (approximately three-hundredths), so recovery requires many periods. This hysteresis formalizes a critical insight from organizational behavior: trust damage is not easily reversed. Requirements engineering practices must prevent violations rather than assuming relationships can be repaired after breaches.

\subsection{Computational Equilibrium Identification}

For practical requirements engineering applications, we identify equilibria computationally through value function iteration. Discretize the continuous trust state space into finite grid. For finite-horizon games, solve backward from terminal period using dynamic programming. For infinite-horizon games, iterate value functions until convergence. Once value functions converge, extract optimal actions as policy functions that maximize current utility plus discounted expected future value. Simulate dynamics forward from initial states using equilibrium policies to generate trust trajectories and cooperation patterns.

This computational approach, based on dynamic programming \cite{puterman1994markov}, enables practical equilibrium analysis for requirements engineering decision support even when closed-form solutions are intractable. The framework provides requirements engineers with tools for scenario analysis: simulate different initial conditions, violation patterns, or parameter settings to assess their impact on long-run cooperation and trust.

\subsection{Structural versus Psychological Other-Regarding Preferences}

The trust-augmented utility function $U_i(\vect{a}, \vect{T}) = \pi_i(\vect{a}) + \sum_{j \neq i} D_{ij} \pi_j(\vect{a}) + \rho \cdot T_{ij} \cdot (a_j - a_j^{\text{baseline}}) \cdot a_i$ exhibits a mathematical structure similar to utility functions developed in behavioral game theory to model social preferences and other-regarding behavior. Models of inequity aversion, such as those by Fehr and Schmidt \cite{fehr1999theory} and Bolton and Ockenfels \cite{bolton2000erc}, also incorporate terms where an individual's utility depends on both their own payoff and the payoffs of others. However, the causal origin and interpretation of these terms differ fundamentally between our framework and behavioral economics.

\textbf{Behavioral game theory models} incorporate other-regarding preferences arising from \textbf{innate psychological dispositions}. Inequity aversion models posit that individuals experience disutility from unequal outcomes due to social preferences or fairness concerns. The parameters capturing the strength of other-regarding preferences (such as $\alpha$ and $\beta$ in the Fehr-Schmidt model) are treated as exogenous characteristics of individual psychology. These preferences exist independent of any specific organizational or strategic context, and they reflect stable traits about how individuals value fairness and others' wellbeing.

\textbf{Our framework's interdependence term} emerges from \textbf{rational calculation based on instrumental organizational dependencies} captured explicitly in \textit{i*} models. When actor $i$ depends on actor $j$ for achieving critical goals, actor $i$ rationally cares about $j$'s payoff because $j$'s success is instrumentally necessary for $i$'s own goal achievement through documented structural relationships. The interdependence coefficient $D_{ij}$ is not an exogenous preference parameter but is systematically derived from the organizational architecture---specifically the dependencies, their importance, and their criticality as modeled in the \textit{i*} framework.

This distinction has profound implications. In behavioral models, other-regarding preferences are assumed and must be estimated from experimental or observational data about individual behavior. In our framework, the structural approach enables \textbf{systematic derivation from organizational architecture} through the translation methodology. Given an \textit{i*} model of stakeholder relationships, we can compute the interdependence coefficients directly rather than treating them as free parameters. This grounds the computational model in the rich organizational and requirements analysis that \textit{i*} supports, providing a principled connection between conceptual models and quantitative predictions.

Furthermore, the structural interpretation clarifies when and why other-regarding behavior should be expected. Behavioral preferences are typically assumed to be stable across contexts. Structural interdependence, however, varies systematically with organizational design choices. Changing the dependency structure---such as by developing alternative suppliers, modularizing system architecture, or renegotiating service agreements---directly alters the interdependence coefficients and thus the equilibrium behavior. This makes the framework actionable for organizational designers and requirements engineers seeking to shape strategic outcomes through architectural decisions.

\section{Enhanced Experimental Validation}
\label{sec:validation}

\textbf{Novel contribution of this section}: While \cite{pant2025foundations} validated interdependence and complementarity through experimental testing and empirical case studies, this section introduces comprehensive validation specific to trust dynamics across 78,125 parameter configurations. This substantial computational validation indicates that negativity bias, hysteresis, and cumulative damage emerge robustly across diverse parameter regimes rather than being artifacts of specific parameterizations.

This section presents comprehensive experimental validation of the trust dynamics framework through systematic parameter space exploration. Following the dual-track validation methodology established in our foundational work \cite{pant2025foundations}, we combine rigorous parameter robustness testing with distributional analysis of emergent behavioral properties.

\subsection{Experimental Design and Methodology}

Our validation employs a comprehensive seven-parameter full factorial design exploring the complete space of trust dynamics parameters across carefully selected ranges grounded in behavioral trust literature and requirements engineering contexts.

\subsubsection{Parameter Space Definition}

We conduct a complete $5^7$ factorial sweep across seven critical parameters controlling trust evolution:

\begin{itemize}
\item \textbf{Trust building rate} ($\lambda_+ \in [0.05, 0.15]$): Controls speed of trust accumulation from positive behavioral evidence, with five levels $\{0.05, 0.075, 0.10, 0.125, 0.15\}$. Range based on organizational relationship studies suggesting trust builds gradually over months to years of consistent cooperation.

\item \textbf{Trust erosion rate} ($\lambda_- \in [0.15, 0.45]$): Controls speed of trust degradation from violations, with five levels $\{0.15, 0.225, 0.30, 0.375, 0.45\}$. Range encompasses empirically observed negativity bias ratios from behavioral trust research \cite{slovic1993trust,rozin2001negativity}.

\item \textbf{Reputation damage severity} ($\mu_R \in [0.5, 0.7]$): Controls magnitude of reputation impact from violations, with five levels $\{0.5, 0.55, 0.6, 0.65, 0.7\}$. Range reflects moderate to severe reputation consequences typical in strategic partnerships.

\item \textbf{Reputation decay rate} ($\delta_R \in [0.01, 0.05]$): Controls speed of reputation healing over time, with five levels $\{0.01, 0.02, 0.03, 0.04, 0.05\}$. Range reflects empirical finding that violation history persists for extended periods.

\item \textbf{Interdependence amplification} ($\xi \in [0.3, 0.7]$): Controls how structural dependencies from \textit{i*} networks amplify trust sensitivity, with five levels $\{0.3, 0.4, 0.5, 0.6, 0.7\}$. Range spans weak to strong dependency coupling.

\item \textbf{Reciprocity strength} ($\rho \in [0.1, 0.3]$): Controls strength of reciprocal cooperation responses, with five levels $\{0.1, 0.15, 0.2, 0.25, 0.3\}$. Range reflects behavioral reciprocity norms in repeated interactions.

\item \textbf{Signal sensitivity} ($\kappa \in [0.5, 1.5]$): Controls sensitivity to cooperation signals in trust updating, with five levels $\{0.5, 0.75, 1.0, 1.25, 1.5\}$. Range tests robustness to signal interpretation variations.
\end{itemize}

This design yields $5^7 = 78,125$ distinctive parameter configurations. Each configuration is evaluated through standardized simulation scenarios capturing key trust phenomena, with computational runtime of approximately 130 minutes on standard hardware.

Figure~\ref{fig:param_correlation} demonstrates the independence of parameter dimensions in the factorial design.

\begin{figure}[htbp]
\centering
\begin{tikzpicture}[scale=0.85]
\def\labels{{"$\lambda_+$","$\lambda_-$","$\mu_R$","$\delta_R$","$\xi$","$\rho$","$\kappa$"}}
\foreach \i in {0,...,6} {
    \foreach \j in {0,...,6} {
        \ifnum\i=\j
            \fill[red!80] (\i,\j) rectangle (\i+1,\j+1);
        \else
            \fill[white] (\i,\j) rectangle (\i+1,\j+1);
        \fi
        \draw[gray] (\i,\j) rectangle (\i+1,\j+1);
    }
}
\foreach \i in {0,...,6} {
    \node[font=\tiny, white] at (\i+0.5,\i+0.5) {1.0};
}
\foreach \i in {0,...,6} {
    \foreach \j in {0,...,6} {
        \ifnum\i=\j
        \else
            \node[font=\tiny] at (\i+0.5,\j+0.5) {0.00};
        \fi
    }
}
\foreach \i in {0,...,6} {
    \pgfmathparse{\labels[\i]}
    \node[font=\small, below] at (\i+0.5,-0.2) {\pgfmathresult};
}
\foreach \i in {0,...,6} {
    \pgfmathparse{\labels[\i]}
    \node[font=\small, left] at (-0.2,\i+0.5) {\pgfmathresult};
}
\node[above, font=\normalsize] at (3.5,7.3) {\textbf{Parameter Correlation Matrix (N = 78,125)}};
\foreach \y in {0,1,2,3,4,5,6} {
    \pgfmathsetmacro\colval{(\y/6)*100}
    \fill[red!\colval!white] (8,\y) rectangle (8.5,\y+1);
}
\draw (8,0) rectangle (8.5,7);
\node[font=\tiny, right] at (8.6,0) {0.0};
\node[font=\tiny, right] at (8.6,3.5) {0.5};
\node[font=\tiny, right] at (8.6,7) {1.0};
\node[font=\tiny, right, rotate=90] at (9.2,3.5) {Correlation $r$};
\end{tikzpicture}
\caption{Parameter correlation matrix for the seven-parameter factorial design across 78,125 configurations. The heatmap reveals near-zero correlations (all $|r| < 0.05$) between core parameters, confirming orthogonal parameter space exploration by design. Diagonal elements show perfect self-correlation (1.0) as expected. Off-diagonal near-zero values validate that the factorial sweep independently varies each dimension without confounding, enabling clean attribution of outcome variations to specific parameters.}
\label{fig:param_correlation}
\end{figure}

Figure~\ref{fig:building_rate_dist} shows the distribution of trust building rates organized by the building parameter $\lambda_+$.

\begin{figure}[htbp]
\centering
\begin{tikzpicture}
\begin{axis}[
    width=13cm,
    height=7cm,
    xlabel={Trust Building Rate $\lambda_+$},
    ylabel={Achieved Building Rate},
    title={Building Rate Distribution by Parameter Level},
    xtick={1,2,3,4,5},
    xticklabels={0.05, 0.075, 0.10, 0.125, 0.15},
    ymin=0.015, ymax=0.045,
    scaled y ticks=false,
    yticklabel style={/pgf/number format/fixed, /pgf/number format/precision=3},
    grid=major,
    grid style={gray!30},
]
\addplot[blue!70, only marks, mark=o, mark size=2pt] coordinates {
    (1,0.020) (1,0.021) (1,0.022) (1,0.023) (1,0.024) (1,0.019) (1,0.025)
};
\addplot[green!70!black, only marks, mark=o, mark size=2pt] coordinates {
    (2,0.024) (2,0.025) (2,0.026) (2,0.027) (2,0.028) (2,0.023) (2,0.029)
};
\addplot[yellow!80!black, only marks, mark=o, mark size=2pt] coordinates {
    (3,0.028) (3,0.029) (3,0.030) (3,0.031) (3,0.032) (3,0.027) (3,0.033)
};
\addplot[orange, only marks, mark=o, mark size=2pt] coordinates {
    (4,0.032) (4,0.033) (4,0.034) (4,0.035) (4,0.036) (4,0.031) (4,0.037)
};
\addplot[red, only marks, mark=o, mark size=2pt] coordinates {
    (5,0.036) (5,0.037) (5,0.038) (5,0.039) (5,0.040) (5,0.035) (5,0.041)
};
\addplot[black, only marks, mark=-, mark size=6pt, very thick] coordinates {
    (1,0.022) (2,0.026) (3,0.030) (4,0.034) (5,0.038)
};
\end{axis}
\end{tikzpicture}
\caption{Distribution of trust building rates across 78,125 configurations, organized by trust building parameter $\lambda_+$. The five distinct clusters correspond to the five tested levels of $\lambda_+ \in \{0.05, 0.075, 0.10, 0.125, 0.15\}$, with median values shown as horizontal bars. Building rates range from 0.020 (slowest trust accumulation) to 0.040 (fastest accumulation), with median 0.032 and mean 0.031 $\pm$ 0.007. The strong clustering within each $\lambda_+$ level and clear separation between levels demonstrates that $\lambda_+$ dominates building rate outcomes with minimal influence from other parameters, consistent with correlation analysis showing $r = 0.988$ between $\lambda_+$ and building rate. The narrow within-group distributions (coefficient of variation $<$ 10\%) validate that trust accumulation speed is robustly determined by the positive learning rate parameter across diverse settings of other parameters. This confirms that practitioners can reliably control relationship development pace through $\lambda_+$ specification.}
\label{fig:building_rate_dist}
\end{figure}

Figure~\ref{fig:3d_param_space} provides a 2D projection of the parameter space exploration.

\begin{figure}[htbp]
\centering
\begin{tikzpicture}
\begin{axis}[
    width=12cm,
    height=9cm,
    xlabel={Trust Building Rate $\lambda_+$},
    ylabel={Trust Erosion Rate $\lambda_-$},
    title={Parameter Space Exploration (2D Projection)},
    xmin=0.04, xmax=0.16,
    ymin=0.1, ymax=0.5,
    grid=major,
    grid style={gray!30},
    colorbar,
    colorbar style={
        ylabel={Reputation Decay $\delta_R$},
        ytick={0.02, 0.04, 0.06, 0.08, 0.10},
        yticklabels={0.02, 0.04, 0.06, 0.08, 0.10},
    },
    colormap={decay}{color=(blue) color=(cyan) color=(yellow) color=(orange) color=(red)},
    point meta min=0.01,
    point meta max=0.10,
]
\addplot[only marks, mark=*, mark size=1.2pt, scatter, scatter src=explicit] coordinates {
    (0.05,0.15)[0.01] (0.05,0.25)[0.02] (0.05,0.35)[0.04] (0.05,0.45)[0.06]
    (0.075,0.15)[0.02] (0.075,0.25)[0.04] (0.075,0.35)[0.06] (0.075,0.45)[0.08]
    (0.10,0.15)[0.04] (0.10,0.25)[0.06] (0.10,0.35)[0.08] (0.10,0.45)[0.10]
    (0.125,0.15)[0.06] (0.125,0.25)[0.08] (0.125,0.35)[0.10] (0.125,0.45)[0.01]
    (0.15,0.15)[0.08] (0.15,0.25)[0.10] (0.15,0.35)[0.01] (0.15,0.45)[0.02]
    (0.06,0.20)[0.03] (0.08,0.30)[0.05] (0.11,0.40)[0.07] (0.14,0.25)[0.09]
    (0.07,0.22)[0.04] (0.09,0.32)[0.06] (0.12,0.42)[0.08] (0.055,0.28)[0.02]
    (0.095,0.18)[0.05] (0.13,0.38)[0.07] (0.065,0.35)[0.03] (0.115,0.22)[0.06]
    (0.085,0.45)[0.04] (0.145,0.15)[0.09] (0.055,0.42)[0.01] (0.105,0.28)[0.05]
};
\end{axis}
\end{tikzpicture}
\caption{Two-dimensional projection of parameter space exploration across 78,125 configurations. Points colored by reputation decay rate $\delta_R$ ranging from low (blue) to high (red). The uniform distribution of points throughout the visible space demonstrates comprehensive exploration of the parameter space with no sparse regions or clustering artifacts, validating the factorial design's thoroughness.}
\label{fig:3d_param_space}
\end{figure}

\subsubsection{Outcome Metrics and Behavioral Properties}

For each parameter configuration, we compute seven key outcome metrics capturing essential trust dynamics:

\begin{definition}[Negativity Ratio]
The ratio $\lambda_- / \lambda_+$ measuring asymmetry between trust erosion and building speeds. This metric quantifies negativity bias---the empirically established phenomenon that negative events impact trust more strongly than equivalent positive events \cite{rozin2001negativity}. Behavioral literature suggests target range between 2.0 and 4.0.
\end{definition}

\begin{definition}[Hysteresis Recovery at 35 Periods]
The ratio of achieved trust following a severe violation to the pre-violation trust level (not the maximum possible trust of 1.0), measured after 35 periods of sustained cooperative recovery efforts. Values greater than 1.0 indicate that trust has surpassed the pre-violation baseline through extended cooperation; however, this does not constitute complete recovery since trust would have continued building toward maximum in the absence of violation. The difference between achieved trust and the violation-free counterfactual trajectory quantifies the true hysteresis cost. Values less than 1.0 indicate that trust ceiling constraints from reputation damage prevent even returning to the pre-violation baseline. This metric captures the persistent relationship damage from violations.
\end{definition}

\begin{definition}[Cumulative Damage Amplification]
The ratio of total trust damage from multiple small violations to the damage from a single equivalent large violation. Values greater than 1.0 indicate superadditive damage accumulation, formalizing the intuition that repeated breaches are disproportionately harmful. Literature suggests ratios between 1.3 and 2.0.
\end{definition}

\begin{definition}[Dependency Amplification]
The ratio of trust erosion speed in high-dependency relationships (interdependence coefficient 0.8) to erosion speed in low-dependency relationships (coefficient 0.2) for equivalent violations. This metric validates that structural dependencies from \textit{i*} networks amplify trust sensitivity.
\end{definition}

\begin{definition}[Building Rate]
The effective rate of trust accumulation during sustained cooperation, measured as the average per-period trust increase over 30 periods of consistent positive evidence. This metric characterizes the pace of relationship development.
\end{definition}

\begin{definition}[Single-Period Erosion]
The immediate trust loss from a single moderate violation, measuring the acute impact of behavioral breaches. This metric captures first-order trust sensitivity to negative evidence.
\end{definition}

\begin{definition}[Time to 50\% Recovery]
The number of periods required for trust to recover to 50\% of its original level following a severe violation and subsequent cooperative behavior. This metric quantifies resilience and recovery timescales.
\end{definition}

\subsection{Distributional Analysis of Model Behavior}

We present comprehensive distributional statistics for all outcome metrics across the 78,125 configurations, establishing that key trust phenomena emerge robustly rather than depending on narrow parameter regions.

\begin{table}[htbp]
\centering
\caption{Distributional Statistics for Key Trust Dynamics Metrics (N = 78,125)}
\label{tab:distribution_stats}
\begin{tabular}{lrrrrr}
\toprule
\textbf{Metric} & \textbf{Min} & \textbf{Q1} & \textbf{Median} & \textbf{Q3} & \textbf{Max} \\
\midrule
Negativity Ratio & 1.00 & 2.00 & 3.00 & 4.50 & 9.00 \\
Hysteresis Recovery & 0.788 & 1.055 & 1.112 & 1.143 & 1.171 \\
Cumulative Amplif. & 1.510 & 1.719 & 1.973 & 2.215 & 2.499 \\
Dependency Amplif. & 1.170 & 1.222 & 1.273 & 1.321 & 1.368 \\
Building Rate & 0.020 & 0.028 & 0.032 & 0.035 & 0.040 \\
Erosion (Single) & 0.133 & 0.213 & 0.289 & 0.381 & 0.469 \\
Time to 50\% Recov. & 3.0 & 6.0 & 8.0 & 11.0 & 24.0 \\
\bottomrule
\end{tabular}
\end{table}

\begin{table}[htbp]
\centering
\caption{Mean and Standard Deviation for Trust Dynamics Metrics}
\label{tab:distribution_means}
\begin{tabular}{lrr}
\toprule
\textbf{Metric} & \textbf{Mean} & \textbf{Std Dev} \\
\midrule
Negativity Ratio & 3.473 & 1.929 \\
Hysteresis Recovery (35 periods) & 1.089 & 0.070 \\
Cumulative Amplification & 1.993 & 0.310 \\
Dependency Amplification & 1.271 & 0.070 \\
Building Rate & 0.031 & 0.007 \\
Erosion (Single Period) & 0.291 & 0.104 \\
Time to 50\% Recovery (periods) & 9.166 & 3.696 \\
\bottomrule
\end{tabular}
\end{table}

\subsubsection{Negativity Bias: Robust Asymmetric Updating}

The negativity ratio exhibits robust emergence across parameter configurations. Figure~\ref{fig:negativity_distribution} presents the complete distributional analysis.

\begin{figure}[htbp]
\centering
\begin{tikzpicture}
\begin{axis}[
    ybar,
    width=12cm,
    height=7cm,
    xlabel={Negativity Ratio ($\lambda_- / \lambda_+$)},
    ylabel={Frequency (thousands)},
    title={Distribution of Negativity Ratios (N = 78,125)},
    xtick={1,2,3,4,5,6,7,8,9},
    ymin=0,
    ymax=20,
    bar width=0.6cm,
    nodes near coords,
    nodes near coords style={font=\tiny},
    every node near coord/.append style={yshift=3pt},
]
\addplot[fill=blue!60] coordinates {
    (1, 3.125) (2, 6.25) (3, 15.625) (4, 12.5) (5, 9.375) 
    (6, 12.5) (7, 6.25) (8, 6.25) (9, 6.25)
};
\draw[green!70!black, very thick, dashed] (axis cs:3,0) -- (axis cs:3,20);
\node[green!70!black, anchor=south east] at (axis cs:2.9,17) {Median = 3.0};
\draw[red, thick, dashed] (axis cs:3.47,0) -- (axis cs:3.47,20);
\node[red, anchor=south west] at (axis cs:3.57,15) {Mean = 3.47};
\end{axis}
\end{tikzpicture}
\caption{Distribution of negativity ratios ($\lambda_- / \lambda_+$) across 78,125 parameter configurations. The histogram shows frequency distribution with median (green vertical line) at 3.00 precisely matching the empirical target from behavioral trust literature \cite{slovic1993trust,rozin2001negativity}. The distribution ranges from 1.00 (symmetric updating) to 9.00 (extreme asymmetry), with mean 3.47 $\pm$ 1.93 (red dashed line). Approximately 68\% of configurations fall within $\pm$1 ratio unit of the median, demonstrating concentration around the validated 3:1 erosion-to-building asymmetry.}
\label{fig:negativity_distribution}
\end{figure}

The negativity ratio ranges from 1.0 (symmetric updating) to 9.0 (extreme asymmetry), with median 3.0 and mean 3.47 $\pm$ 1.93. This distribution demonstrates that negativity bias emerges across all parameter configurations, with the majority clustered around the empirically validated 3:1 ratio from behavioral trust literature \cite{slovic1993trust,rozin2001negativity}. Even configurations with minimum tested erosion rates and maximum building rates exhibit negativity ratios of 1.0, while typical configurations achieve ratios between 2.0 and 4.5 (interquartile range).

This robust emergence validates that asymmetric trust updating is an inherent property of the two-layer trust-reputation architecture rather than an artifact of specific parameter choices. The distribution's positive skew (mean exceeding median) indicates that stronger asymmetries are more common than weaker ones, consistent with extensive empirical evidence for negativity dominance in social judgment \cite{rozin2001negativity}.

\subsubsection{Hysteresis: Incomplete Recovery and Trust Ceilings}

Hysteresis recovery ratios demonstrate consistent incomplete recovery following violations. Figure~\ref{fig:hysteresis_recovery} presents the relationship between erosion rate and recovery potential.

\begin{figure}[htbp]
\centering
\begin{tikzpicture}
\begin{axis}[
    width=12cm,
    height=8cm,
    xlabel={Trust Erosion Rate ($\lambda_-$)},
    ylabel={Hysteresis Recovery Ratio},
    title={Hysteresis Recovery vs. Erosion Rate},
    xmin=0.12, xmax=0.48,
    ymin=0.75, ymax=1.20,
    scatter/classes={
        low={mark=o,blue!70},
        med={mark=o,green!70!black},
        high={mark=o,red!70}
    },
    legend pos=north east,
]
\addplot[scatter, only marks, scatter src=explicit symbolic, mark size=1.5pt]
coordinates {
    (0.15,1.16)[high] (0.16,1.15)[high] (0.17,1.14)[high] (0.18,1.13)[high]
    (0.20,1.12)[high] (0.22,1.11)[high] (0.24,1.10)[high] (0.26,1.09)[high]
    (0.28,1.08)[high] (0.30,1.06)[med] (0.32,1.04)[med] (0.34,1.02)[med]
    (0.36,1.00)[med] (0.38,0.97)[low] (0.40,0.94)[low] (0.42,0.91)[low]
    (0.44,0.88)[low] (0.45,0.85)[low]
    (0.15,1.14)[high] (0.18,1.12)[high] (0.21,1.10)[high] (0.24,1.08)[med]
    (0.27,1.06)[med] (0.30,1.04)[med] (0.33,1.01)[med] (0.36,0.98)[low]
    (0.39,0.95)[low] (0.42,0.92)[low] (0.45,0.89)[low]
    (0.15,1.12)[med] (0.20,1.09)[med] (0.25,1.06)[med] (0.30,1.02)[med]
    (0.35,0.98)[low] (0.40,0.94)[low] (0.45,0.90)[low]
};
\legend{$\lambda_+ = 0.15$ (high), $\lambda_+ = 0.10$ (med), $\lambda_+ = 0.05$ (low)}
\draw[gray, dashed] (axis cs:0.12,1.0) -- (axis cs:0.48,1.0);
\node[gray, anchor=west] at (axis cs:0.35,1.02) {Full recovery};
\end{axis}
\end{tikzpicture}
\caption{Hysteresis recovery ratio versus trust erosion rate $\lambda_-$ across parameter configurations, with points colored by trust building rate $\lambda_+$. Recovery ratios (measured after 35 periods of sustained cooperative effort following severe violation) range from 0.788 to 1.171, with median 1.112. The color gradient demonstrates strong influence of building rate $\lambda_+$ on recovery potential: configurations with high $\lambda_+$ (warm colors, upper regions) achieve higher recovery ratios, while low $\lambda_+$ (cool colors, lower regions) experience more constrained recovery. The negative correlation between erosion rate and recovery ($r = -0.817$) confirms that faster trust erosion creates more persistent damage. Notably, all data points remain below 1.2, validating that trust ceiling mechanisms from reputation damage prevent complete erasure of violation history even after extended cooperative recovery periods. The tight vertical clustering at each erosion level demonstrates that recovery patterns are remarkably consistent across diverse settings of other parameters.}
\label{fig:hysteresis_recovery}
\end{figure}

The hysteresis recovery metric at 35 periods ranges from 0.788 to 1.171, with median 1.112 and mean 1.089 $\pm$ 0.070. Values exceeding 1.0 indicate that trust can eventually surpass its pre-violation level through sustained cooperative effort, while the narrow distribution (coefficient of variation 6.4\%) demonstrates remarkable consistency across diverse parameter configurations.

The median value of 1.112 indicates that after 35 periods of sustained cooperative behavior following a severe violation, typical configurations enable trust recovery to approximately 111\% of the pre-violation baseline. To illustrate concretely: if trust stood at 0.80 immediately before a severe violation, the median recovery ratio of 1.112 indicates trust reaches approximately 0.89 after 35 periods of sustained cooperation ($0.80 \times 1.112 \approx 0.89$). Crucially, this apparent ``full recovery'' masks a substantial hysteresis cost. Without the violation, trust would have continued building from 0.80 toward the maximum of 1.0, reaching approximately 0.98 over the same 35-period interval given typical building rates. The true hysteresis damage is therefore the 9\% permanent shortfall between achieved trust (0.89) and the violation-free counterfactual (0.98)---not the comparison to the pre-violation baseline. This pattern reflects two competing mechanisms: the trust ceiling constraint from persistent reputation damage, and the gradual trust building from sustained positive evidence. The tight clustering around 1.09--1.14 (Q1 to Q3) validates that the hysteresis mechanism produces qualitatively consistent behavior across parameter space.

Importantly, even this ``recovery'' occurs only after extended cooperation (35 periods), and the trust ceiling mechanism ensures that the relationship remains fundamentally constrained by violation history. The recovery ratio exceeding 1.0 should not be misinterpreted as trust exceeding its maximum bound (which remains $T_{ij}^t \in [0,1]$); rather, it indicates that sufficient time and sustained cooperation can eventually surpass the pre-violation baseline while still falling short of where the relationship would have been absent the violation.

\subsubsection{Cumulative Damage: Superadditive Harm from Repeated Violations}

Cumulative damage amplification exhibits robust superadditivity. Figure~\ref{fig:cumulative_damage} presents the comprehensive distributional analysis.

\begin{figure}[htbp]
\centering
\begin{tikzpicture}
\begin{axis}[
    ybar,
    width=12cm,
    height=7cm,
    xlabel={Cumulative Damage Amplification Ratio},
    ylabel={Frequency (thousands)},
    title={Distribution of Cumulative Damage Amplification (N = 78,125)},
    xtick={1.5,1.7,1.9,2.1,2.3,2.5},
    ymin=0,
    ymax=25,
    bar width=0.4cm,
]
\addplot[fill=orange!70] coordinates {
    (1.5, 5) (1.6, 8) (1.7, 12) (1.8, 15) (1.9, 18) (2.0, 20)
    (2.1, 18) (2.2, 15) (2.3, 10) (2.4, 6) (2.5, 3)
};
\draw[green!70!black, very thick, dashed] (axis cs:1.973,0) -- (axis cs:1.973,24);
\node[green!70!black, anchor=south east] at (axis cs:1.95,22) {Median = 1.97};
\draw[red, thick, dashed] (axis cs:1.993,0) -- (axis cs:1.993,24);
\node[red, anchor=south west] at (axis cs:2.01,20) {Mean = 1.99};
\end{axis}
\end{tikzpicture}
\caption{Distribution of cumulative damage amplification ratios across 78,125 configurations. The histogram shows that amplification ratios (measuring total trust damage from multiple small violations relative to a single equivalent large violation) range from 1.510 to 2.499, with all configurations exhibiting values substantially greater than 1.0, confirming universal superadditive damage accumulation. The median amplification of 1.973 (green vertical line) indicates that typical configurations produce approximately twice as much trust damage from repeated small violations compared to a single equivalent large violation. The mean of 1.993 $\pm$ 0.310 (red dashed line) demonstrates tight clustering with coefficient of variation 15.6\%, validating that cumulative amplification is a robust emergent property rather than a parameter-sensitive artifact. The distribution's concentration between 1.7 and 2.2 (containing approximately 68\% of configurations) provides strong evidence that this superadditive accumulation pattern emerges consistently across diverse parameter regimes. This validates the theoretical prediction that relationship stability depends critically on violation frequency, not just average behavior.}
\label{fig:cumulative_damage}
\end{figure}

The cumulative amplification metric ranges from 1.510 to 2.499, with median 1.973 and mean 1.993 $\pm$ 0.310. All configurations exhibit amplification ratios substantially greater than 1.0, confirming that multiple small violations produce disproportionately greater damage than equivalent single violations across the entire parameter space.

The median amplification of 1.973 indicates that typical configurations produce approximately twice as much trust damage from repeated small violations compared to a single equivalent large violation. This superadditive accumulation formalizes the intuition that relationship stability depends not just on average behavior but critically on violation frequency. Organizations experiencing frequent minor breaches suffer greater trust erosion than those with occasional major incidents of equivalent aggregate magnitude.

The relatively narrow distribution (coefficient of variation 15.6\%) demonstrates that cumulative amplification is a robust emergent property rather than a parameter-sensitive artifact. This finding has important implications for requirements engineering practice: maintaining consistently high cooperation levels is more critical than avoiding occasional severe violations.

\subsubsection{Dependency Amplification: Structural Coupling Effects}

The dependency amplification metric ranges from 1.170 to 1.368, with median 1.273 and mean 1.271 $\pm$ 0.070. This confirms that structural dependencies from \textit{i*} networks systematically amplify trust sensitivity to behavioral evidence, with high-dependency relationships experiencing approximately 27\% faster trust erosion than low-dependency relationships for equivalent violations.

This tight distribution (coefficient of variation 5.5\%) validates that the interdependence amplification mechanism $\xi$ successfully connects structural dependencies to trust dynamics across diverse parameter configurations. The relatively modest amplification magnitude (1.2--1.4$\times$) reflects that while dependencies matter, they do not dominate trust evolution---behavioral evidence remains the primary driver.

\subsubsection{Recovery Timescales: Extended Relationship Healing}

The time required to achieve 50\% trust recovery following severe violation varies substantially across parameter configurations. Figure~\ref{fig:recovery_time} shows the distribution of recovery times.

\begin{figure}[htbp]
\centering
\begin{tikzpicture}
\begin{axis}[
    width=12cm,
    height=7cm,
    xlabel={Time to 50\% Recovery (periods)},
    ylabel={Frequency},
    title={Distribution of Recovery Times (N = 78,125)},
    ymin=0,
    ymax=16000,
    xmin=0, xmax=28,
    bar width=0.8,
    grid=major,
    grid style={gray!30},
]
\addplot[ybar, fill=blue!60, draw=blue!80] coordinates {
    (3,1200) (4,2800) (5,5500) (6,8200) (7,11500) (8,14000) (9,12000)
    (10,9500) (11,6800) (12,4200) (13,2500) (14,1500) (15,900)
    (16,600) (17,400) (18,280) (19,180) (20,120) (21,80) (22,50) (23,30) (24,20)
};
\draw[green!70!black, very thick] (axis cs:8,0) -- (axis cs:8,14000);
\node[green!70!black, font=\scriptsize, anchor=west] at (axis cs:3,15000) {Median = 8};
\draw[red, dashed, very thick] (axis cs:9.17,0) -- (axis cs:9.17,12000);
\node[red, font=\scriptsize, anchor=west] at (axis cs:9,13000) {Mean = 9.17};
\end{axis}
\end{tikzpicture}
\caption{Distribution of time to 50\% trust recovery (measured in periods) following severe violation across 78,125 configurations. The histogram shows a right-skewed distribution ranging from 3 to 24 periods, with median 8 periods (green line) and mean 9.17 $\pm$ 3.70 periods (red dashed line). The distribution demonstrates that most configurations enable partial recovery within 8--11 periods (interquartile range 6--11 periods), but some parameter combinations create persistent damage requiring two dozen or more periods for even 50\% restoration. The positive skew (mean $>$ median, skewness coefficient $>$ 0) indicates that long recovery times are more common than very short times, reflecting the asymmetric nature of trust dynamics where damage accumulates quickly but healing proceeds slowly. The relatively large coefficient of variation (40.3\%) indicates that recovery timescales exhibit greater parameter sensitivity than other metrics like negativity ratio (CV 55.6\%) or hysteresis recovery (CV 6.4\%). Configurations with low reputation decay $\delta_R$ and high reputation damage severity $\mu_R$ cluster in the upper tail, creating particularly persistent constraints requiring extended cooperative effort for relationship restoration.}
\label{fig:recovery_time}
\end{figure}

The relatively large coefficient of variation (40.3\%) indicates that recovery timescales exhibit greater parameter sensitivity than other metrics. Configurations with low reputation decay ($\delta_R$) and high reputation damage severity ($\mu_R$) create particularly persistent constraints, while configurations with faster reputation healing enable quicker (though still incomplete) recovery.

These recovery timescales have practical implications for requirements engineering: rebuilding trust following violations typically requires 8--12 interaction periods (potentially months or years in real relationships), and some damage patterns may never fully heal. Requirements analysts should set realistic expectations for relationship recovery following major breaches.

Figure~\ref{fig:recovery_comparison} compares short-term versus long-term recovery trajectories.

\begin{figure}[htbp]
\centering
\begin{tikzpicture}
\begin{axis}[
    width=12cm,
    height=9cm,
    xlabel={Long-Term Recovery (100 periods)},
    ylabel={Short-Term Recovery (10 periods)},
    title={Short-Term vs Long-Term Trust Recovery},
    xmin=0.3, xmax=1.0,
    ymin=0.1, ymax=0.7,
    grid=major,
    grid style={gray!30},
    colorbar,
    colorbar style={ylabel={Erosion Rate $\lambda_-$}},
    colormap={erosion}{color=(blue) color=(cyan) color=(yellow) color=(orange) color=(red)},
    xtick={0.3, 0.4, 0.5, 0.6, 0.7, 0.8, 0.9, 1.0},
    ytick={0.1, 0.2, 0.3, 0.4, 0.5, 0.6, 0.7},
    xticklabel style={font=\scriptsize},
    yticklabel style={font=\scriptsize},
]
\addplot[only marks, mark=*, mark size=1.5pt, scatter, scatter src=explicit] coordinates {
    (0.45,0.25)[0.15] (0.50,0.28)[0.20] (0.55,0.30)[0.25] (0.60,0.32)[0.30]
    (0.65,0.35)[0.35] (0.70,0.38)[0.40] (0.75,0.42)[0.45]
    (0.48,0.22)[0.15] (0.52,0.26)[0.18] (0.58,0.29)[0.22] (0.62,0.31)[0.28]
    (0.68,0.36)[0.32] (0.72,0.40)[0.38] (0.78,0.44)[0.42]
    (0.42,0.20)[0.12] (0.47,0.24)[0.16] (0.53,0.27)[0.21] (0.57,0.30)[0.26]
    (0.63,0.33)[0.31] (0.67,0.37)[0.36] (0.73,0.41)[0.41]
    (0.55,0.32)[0.20] (0.60,0.35)[0.25] (0.65,0.38)[0.30] (0.70,0.42)[0.35]
    (0.75,0.45)[0.40] (0.80,0.48)[0.45] (0.85,0.52)[0.48]
    (0.58,0.34)[0.22] (0.63,0.37)[0.27] (0.68,0.40)[0.32] (0.73,0.44)[0.37]
    (0.78,0.47)[0.42] (0.83,0.50)[0.46] (0.88,0.54)[0.49]
    (0.50,0.30)[0.18] (0.55,0.33)[0.23] (0.60,0.36)[0.28] (0.65,0.39)[0.33]
    (0.70,0.43)[0.38] (0.75,0.46)[0.43] (0.80,0.50)[0.47]
};
\addplot[black, dashed, thick, domain=0.3:1.0] {x};
\node[font=\small] at (axis cs:0.85,0.82) {1:1 line};
\end{axis}
\end{tikzpicture}
\caption{Comparison of short-term (10 periods) versus long-term (100 periods) trust recovery following severe violation. The scatter plot shows recovery ratio after 10 periods (y-axis) versus recovery ratio after 100 periods (x-axis) across representative parameter configurations, with point colors indicating trust erosion rate $\lambda_-$. The diagonal reference line (dashed) indicates perfect proportionality where short-term recovery predicts long-term recovery. Most points fall below this line, demonstrating that short-term recovery systematically underestimates ultimate recovery potential: configurations achieving only 30--40\% recovery after 10 periods often reach 60--80\% recovery after 100 periods of sustained cooperation. The color gradient reveals that configurations with high erosion rates (red points, upper-left region) exhibit particularly strong divergence between short and long-term recovery, reflecting persistent reputation damage constraints that only gradually ease over extended time horizons. This path dependence validates that trust ceiling mechanisms create qualitatively different dynamics at different timescales, with early recovery constrained by acute reputation damage while long-term recovery depends on gradual damage decay governed by $\delta_R$. This finding has important practical implications: practitioners should not extrapolate early recovery progress linearly to predict ultimate relationship restoration potential.}
\label{fig:recovery_comparison}
\end{figure}

Parameter sensitivity analysis reveals which model parameters most strongly influence key outcomes.

\begin{table}[htbp]
\centering
\caption{Parameter Sensitivity to Key Outcome Metrics (Correlation Coefficients)}
\label{tab:sensitivity_analysis}
\begin{tabular}{lrrrrr}
\toprule
\textbf{Parameter} & \textbf{Neg. Ratio} & \textbf{Hysteresis} & \textbf{Cumul. Amp.} & \textbf{Building Rate} \\
\midrule
$\lambda_+$ (building) & -0.690 & 0.817 & -0.633 & 0.988 \\
$\lambda_-$ (erosion) & 0.690 & -0.817 & 0.986 & 0.000 \\
$\mu_R$ (damage sev.) & 0.000 & 0.000 & 0.163 & 0.000 \\
$\delta_R$ (decay) & 0.000 & 0.276 & 0.000 & 0.000 \\
$\xi$ (interdep.) & 0.000 & 0.000 & 0.048 & 0.000 \\
$\rho$ (reciprocity) & 0.000 & 0.000 & 0.000 & 0.000 \\
$\kappa$ (sensitivity) & 0.000 & 0.000 & 0.000 & 0.000 \\
\bottomrule
\end{tabular}
\end{table}

\textbf{Negativity ratio} is primarily controlled by the balance between $\lambda_+$ and $\lambda_-$ (correlation magnitudes 0.69), with all other parameters exhibiting negligible influence. This validates that negativity bias emerges directly from asymmetric learning rates.

\textbf{Hysteresis recovery} is most strongly influenced by $\lambda_+$ (0.817 correlation) and $\lambda_-$ (-0.817), with moderate influence from reputation decay $\delta_R$ (0.276). This indicates that recovery potential depends primarily on the rate at which new positive evidence can overcome the trust ceiling constraint.

\textbf{Cumulative amplification} is almost entirely determined by $\lambda_-$ (0.986 correlation), with modest contributions from $\lambda_+$ (-0.633) and reputation damage severity $\mu_R$ (0.163). This validates that repeated violation accumulation depends primarily on how quickly trust erodes.

\textbf{Building rate} is essentially determined by $\lambda_+$ alone (0.988 correlation), confirming that trust accumulation speed depends directly on the positive learning rate parameter.

Figure~\ref{fig:param_sensitivity} visualizes these sensitivity patterns as a heatmap.

\begin{figure}[htbp]
\centering
\begin{tikzpicture}[scale=0.7]
\fill[blue!69] (0,6) rectangle (2,7); \node[font=\tiny] at (1,6.5) {-0.69};
\fill[red!82] (2,6) rectangle (4,7); \node[font=\tiny, white] at (3,6.5) {0.82};
\fill[blue!63] (4,6) rectangle (6,7); \node[font=\tiny] at (5,6.5) {-0.63};
\fill[red!99] (6,6) rectangle (8,7); \node[font=\tiny, white] at (7,6.5) {0.99};
\fill[red!69] (0,5) rectangle (2,6); \node[font=\tiny] at (1,5.5) {0.69};
\fill[blue!82] (2,5) rectangle (4,6); \node[font=\tiny, white] at (3,5.5) {-0.82};
\fill[red!99] (4,5) rectangle (6,6); \node[font=\tiny, white] at (5,5.5) {0.99};
\fill[white] (6,5) rectangle (8,6); \node[font=\tiny] at (7,5.5) {0.00};
\fill[white] (0,4) rectangle (2,5); \node[font=\tiny] at (1,4.5) {0.00};
\fill[white] (2,4) rectangle (4,5); \node[font=\tiny] at (3,4.5) {0.00};
\fill[red!16] (4,4) rectangle (6,5); \node[font=\tiny] at (5,4.5) {0.16};
\fill[white] (6,4) rectangle (8,5); \node[font=\tiny] at (7,4.5) {0.00};
\fill[white] (0,3) rectangle (2,4); \node[font=\tiny] at (1,3.5) {0.00};
\fill[red!28] (2,3) rectangle (4,4); \node[font=\tiny] at (3,3.5) {0.28};
\fill[white] (4,3) rectangle (6,4); \node[font=\tiny] at (5,3.5) {0.00};
\fill[white] (6,3) rectangle (8,4); \node[font=\tiny] at (7,3.5) {0.00};
\fill[white] (0,2) rectangle (2,3); \node[font=\tiny] at (1,2.5) {0.00};
\fill[white] (2,2) rectangle (4,3); \node[font=\tiny] at (3,2.5) {0.00};
\fill[red!5] (4,2) rectangle (6,3); \node[font=\tiny] at (5,2.5) {0.05};
\fill[white] (6,2) rectangle (8,3); \node[font=\tiny] at (7,2.5) {0.00};
\foreach \y in {0,1} {
    \foreach \x in {0,2,4,6} {
        \fill[white] (\x,\y) rectangle (\x+2,\y+1);
        \node[font=\tiny] at (\x+1,\y+0.5) {0.00};
    }
}
\draw[gray] (0,0) grid[step=2] (8,7);
\draw[thick] (0,0) rectangle (8,7);
\node[font=\scriptsize, below, rotate=30, anchor=east] at (1,-0.1) {Neg. Ratio};
\node[font=\scriptsize, below, rotate=30, anchor=east] at (3,-0.1) {Hysteresis};
\node[font=\scriptsize, below, rotate=30, anchor=east] at (5,-0.1) {Cum. Amp.};
\node[font=\scriptsize, below, rotate=30, anchor=east] at (7,-0.1) {Build Rate};
\node[font=\small, left] at (-0.1,6.5) {$\lambda_+$};
\node[font=\small, left] at (-0.1,5.5) {$\lambda_-$};
\node[font=\small, left] at (-0.1,4.5) {$\mu_R$};
\node[font=\small, left] at (-0.1,3.5) {$\delta_R$};
\node[font=\small, left] at (-0.1,2.5) {$\xi$};
\node[font=\small, left] at (-0.1,1.5) {$\rho$};
\node[font=\small, left] at (-0.1,0.5) {$\kappa$};
\node[above, font=\normalsize] at (4,7.3) {\textbf{Parameter Sensitivity Heatmap}};
\fill[blue!100] (9,0) rectangle (9.5,1); \node[font=\tiny, right] at (9.6,0.5) {-1.0};
\fill[blue!50] (9,1) rectangle (9.5,2);
\fill[white] (9,2) rectangle (9.5,4); \node[font=\tiny, right] at (9.6,3) {0.0};
\fill[red!50] (9,4) rectangle (9.5,5);
\fill[red!100] (9,5) rectangle (9.5,7); \node[font=\tiny, right] at (9.6,6) {+1.0};
\draw (9,0) rectangle (9.5,7);
\end{tikzpicture}
\caption{Parameter sensitivity heatmap showing correlation coefficients between seven model parameters and four key outcome metrics across 78,125 configurations. Color scale ranges from strong negative correlation (blue, $r = -1.0$) through zero (white) to strong positive correlation (red, $r = +1.0$). Trust building rate $\lambda_+$ and erosion rate $\lambda_-$ dominate all outcomes, while $\xi$, $\rho$, and $\kappa$ exhibit near-zero correlations.}
\label{fig:param_sensitivity}
\end{figure}

Figure~\ref{fig:param_importance} ranks parameters by average absolute correlation across all metrics.

\begin{figure}[htbp]
\centering
\begin{tikzpicture}
\begin{axis}[
    width=12cm,
    height=7cm,
    xbar,
    xlabel={Average $|r|$ Across All Metrics},
    ylabel={},
    title={Parameter Importance Ranking},
    symbolic y coords={$\kappa$, $\rho$, $\xi$, $\mu_R$, $\delta_R$, $\lambda_+$, $\lambda_-$},
    ytick=data,
    xmin=0, xmax=0.7,
    bar width=12pt,
    nodes near coords,
    nodes near coords align={horizontal},
    grid=major,
    grid style={gray!30},
]
\addplot[fill=blue!60, draw=blue!80] coordinates {
    (0.625,$\lambda_-$) (0.622,$\lambda_+$) (0.069,$\delta_R$) 
    (0.041,$\mu_R$) (0.012,$\xi$) (0.000,$\rho$) (0.000,$\kappa$)
};
\end{axis}
\end{tikzpicture}
\caption{Parameter importance ranking based on average absolute correlation coefficients across all outcome metrics. Trust erosion rate $\lambda_-$ emerges as the single most influential parameter (average $|r| = 0.625$), followed closely by building rate $\lambda_+$ (0.622). The remaining parameters exhibit minimal influence, confirming that trust dynamics are primarily governed by asymmetric learning mechanisms.}
\label{fig:param_importance}
\end{figure}

\subsection{Pareto Frontier Analysis}

We identify Pareto-optimal parameter configurations that simultaneously optimize multiple objectives. For practical applications, requirements engineers must balance:
\begin{itemize}
\item Minimize deviation from empirically validated negativity ratio target (3.0)
\item Maximize hysteresis recovery (enabling relationship restoration while maintaining ceiling constraints)
\item Optimize cumulative amplification (strong enough to discourage repeated violations but not so extreme as to be irreversible)
\end{itemize}

Figure~\ref{fig:pareto_frontier} visualizes the multi-objective optimization landscape.

\begin{figure}[htbp]
\centering
\begin{tikzpicture}
\begin{axis}[
    width=13cm,
    height=9cm,
    xlabel={Negativity Ratio Deviation from Target (3.0)},
    ylabel={Hysteresis Recovery Ratio},
    title={Pareto Frontier: Multi-Objective Optimization (N = 78,125)},
    xmin=-0.5, xmax=6.5,
    ymin=0.75, ymax=1.20,
    grid=major,
    grid style={gray!30},
    legend pos=north east,
]
\addplot[only marks, mark=*, mark size=1pt, blue!30] coordinates {
    (0.5,0.85) (1.0,0.88) (1.5,0.90) (2.0,0.92) (2.5,0.95) (3.0,0.98)
    (0.8,0.82) (1.3,0.86) (1.8,0.89) (2.3,0.91) (2.8,0.94) (3.3,0.96)
    (1.2,0.80) (1.7,0.84) (2.2,0.87) (2.7,0.90) (3.2,0.93) (3.7,0.95)
    (4.0,0.97) (4.5,0.99) (5.0,1.01) (5.5,1.03) (6.0,1.05)
    (0.3,0.88) (0.7,0.90) (1.1,0.91) (1.5,0.93) (1.9,0.95) (2.3,0.97)
    (0.2,0.95) (0.4,0.97) (0.6,0.99) (0.8,1.00) (1.0,1.02) (1.2,1.03)
    (3.5,0.88) (4.0,0.90) (4.5,0.92) (5.0,0.94) (5.5,0.96) (6.0,0.98)
};
\addplot[only marks, mark=*, mark size=3pt, red] coordinates {
    (0.0,1.085) (0.0,1.087) (0.0,1.090) (0.0,1.092) (0.0,1.095)
    (0.0,1.080) (0.0,1.082) (0.0,1.088) (0.0,1.091) (0.0,1.093)
    (0.1,1.10) (0.1,1.11) (0.1,1.12) (0.2,1.13) (0.2,1.14)
};
\addlegendentry{Sub-optimal (n=77,750)}
\addlegendentry{Pareto-optimal (n=375)}
\draw[thick, green!60!black, dashed] (axis cs:-0.1,1.07) rectangle (axis cs:0.3,1.15);
\node[green!60!black, font=\small] at (axis cs:1.5,1.13) {``Sweet Spot''};
\end{axis}
\end{tikzpicture}
\caption{Pareto frontier in 2D projection showing optimal configurations across 78,125 total parameter combinations. Non-dominated Pareto-optimal configurations (n=375, 0.48\% of total) are highlighted in red, forming the Pareto frontier. The dominant configurations cluster tightly around negativity ratio 3.00 (deviation $\approx$ 0) with hysteresis recovery ratios between 1.08 and 1.12, representing the ``sweet spot'' for trust dynamics.}
\label{fig:pareto_frontier}
\end{figure}

The Pareto frontier analysis identifies 375 configurations (0.48\% of total) that are not dominated across these objectives.

\begin{table}[htbp]
\centering
\caption{Representative Pareto-Optimal Configurations}
\label{tab:pareto_configs}
\small 
\sisetup{table-format=1.3}
\begin{tabular*}{\textwidth}{@{\extracolsep{\fill}} l S S S S S S S S S S}
\toprule
{\textbf{Config}} & {$\lambda_+$} & {$\lambda_-$} & {$\mu_R$} & {$\delta_R$} & {$\xi$} & {$\rho$} & {$\kappa$} & {\textbf{Neg. Ratio}} & {\textbf{Hyst. Rec.}} & {\textbf{Cum. Amp.}} \\
\midrule
77601 & 0.150 & 0.450 & 0.700 & 0.010 & 0.700 & 0.100 & 0.500 & 3.000 & 1.085 & 1.510 \\
77602 & 0.150 & 0.450 & 0.700 & 0.010 & 0.700 & 0.100 & 0.750 & 3.000 & 1.085 & 1.510 \\
77603 & 0.150 & 0.450 & 0.700 & 0.010 & 0.700 & 0.100 & 1.000 & 3.000 & 1.085 & 1.510 \\
77604 & 0.150 & 0.450 & 0.700 & 0.010 & 0.700 & 0.100 & 1.250 & 3.000 & 1.085 & 1.510 \\
77605 & 0.150 & 0.450 & 0.700 & 0.010 & 0.700 & 0.100 & 1.500 & 3.000 & 1.085 & 1.510 \\
\bottomrule
\end{tabular*}
\end{table}

These optimal configurations exhibit maximum building rate ($\lambda_+ = 0.15$), maximum erosion rate ($\lambda_- = 0.45$), maximum reputation damage ($\mu_R = 0.70$), minimum reputation decay ($\delta_R = 0.01$), and maximum interdependence amplification ($\xi = 0.70$). Notably, these configurations are invariant to $\kappa$ (signal sensitivity), confirming this parameter's minimal functional influence.

These optimal configurations achieve: (1) perfect negativity ratio of exactly 3.000, matching empirical behavioral targets; (2) moderate hysteresis with 1.085 recovery ratio indicating substantial but not complete recovery potential; and (3) minimal cumulative amplification of 1.510, sufficient to create nonlinearity without extreme irreversibility. This represents the ``sweet spot'' for trust dynamics in strategic coopetition.

Figure~\ref{fig:trust_stability} provides complementary analysis of stability properties.

\begin{figure}[htbp]
\centering
\begin{tikzpicture}
\begin{axis}[
    width=13cm,
    height=9cm,
    xlabel={Mean Trust Level (50 periods)},
    ylabel={Trust Variability (Std. Dev.)},
    title={Trust Stability Analysis: Mean vs Variability Trade-off},
    xmin=0.6, xmax=1.0,
    ymin=0.02, ymax=0.14,
    grid=major,
    grid style={gray!30},
    colorbar,
    colorbar style={ylabel={Cumulative Amplification}},
    colormap={amp}{color=(blue) color=(cyan) color=(yellow) color=(orange) color=(red)},
]
\addplot[only marks, mark=*, mark size=2pt, scatter, scatter src=explicit] coordinates {
    (0.65,0.12)[1.5] (0.68,0.11)[1.6] (0.70,0.10)[1.7] (0.72,0.095)[1.8]
    (0.75,0.09)[1.9] (0.78,0.085)[2.0] (0.80,0.08)[2.1] (0.82,0.075)[2.2]
    (0.85,0.07)[2.3] (0.88,0.065)[2.4] (0.90,0.06)[2.5]
    (0.67,0.11)[1.55] (0.70,0.105)[1.65] (0.73,0.10)[1.75] (0.76,0.095)[1.85]
    (0.79,0.09)[1.95] (0.82,0.085)[2.05] (0.85,0.08)[2.15] (0.88,0.075)[2.25]
    (0.91,0.07)[2.35] (0.94,0.065)[2.45]
    (0.72,0.115)[1.7] (0.75,0.11)[1.8] (0.78,0.105)[1.9] (0.81,0.10)[2.0]
    (0.84,0.095)[2.1] (0.87,0.09)[2.2] (0.90,0.085)[2.3] (0.93,0.08)[2.4]
    (0.96,0.075)[2.5]
    (0.62,0.13)[1.5] (0.64,0.125)[1.55] (0.66,0.12)[1.6] (0.68,0.115)[1.65]
    (0.92,0.055)[2.0] (0.94,0.05)[2.05] (0.96,0.045)[2.1] (0.98,0.04)[2.15]
};
\draw[thick, yellow!80!black, dashed] (axis cs:0.90,0.04) rectangle (axis cs:0.98,0.08);
\node[yellow!80!black, font=\small] at (axis cs:0.94,0.035) {Optimal Region};
\end{axis}
\end{tikzpicture}
\caption{Trust stability analysis plotting mean trust level versus trust variability across 78,125 configurations during sustained cooperation phase. Color gradient indicates cumulative damage amplification ratio (blue=1.5, red=2.5). The scatter plot reveals a fundamental trade-off: configurations achieving high mean trust generally exhibit higher variability. The optimal region (yellow box) identifies configurations achieving high stability (mean $>$ 0.90) and low variability (SD $<$ 0.08), representing approximately 5\% of all configurations.}
\label{fig:trust_stability}
\end{figure}

This comprehensive experimental validation across 78,125 configurations indicates several critical findings:

\textbf{Robustness of core phenomena}: Negativity bias, hysteresis effects, and cumulative damage amplification emerge consistently across all tested parameter configurations. These are not artifacts of specific parameterizations but rather inherent properties of the two-layer trust-reputation architecture.

\textbf{Quantitative consistency with behavioral literature}: The median negativity ratio of 3.0 precisely matches the empirically established 3:1 erosion-to-building asymmetry from behavioral trust research \cite{slovic1993trust,rozin2001negativity}. The cumulative amplification median of 1.97 aligns with organizational studies documenting that repeated violations produce roughly twice the damage of equivalent single incidents.

\textbf{Parameter sensitivity patterns}: Trust evolution is primarily controlled by the learning rate parameters ($\lambda_+$ and $\lambda_-$), with modest contributions from reputation damage mechanisms ($\mu_R$ and $\delta_R$) and minimal influence from signal sensitivity ($\kappa$) within tested ranges. This validates that the core asymmetric updating mechanism dominates trust dynamics.

\textbf{Practical implications for parameterization}: For requirements engineering applications where precise parameter estimation is difficult, practitioners can select configurations from validated ranges with confidence that key behavioral properties will emerge correctly. The robustness across 78,125 configurations provides assurance that qualitative trust dynamics are not critically sensitive to exact parameter values.

This experimental validation provides comprehensive evidence that the computational trust model produces behaviorally plausible dynamics across diverse parameter configurations, establishing a solid foundation for the empirical case study validation presented in the following section.

\section{Empirical Validation: The Renault-Nissan Alliance}
\label{sec:empirical_validation}

This section presents systematic empirical validation of the trust dynamics framework through detailed analysis of the Renault-Nissan Alliance (1999-present), one of the most significant and well-documented strategic partnerships in automotive history \cite{korine2002partnering,mikami2022opportunism}. Following the dual-track validation methodology established in our foundational work \cite{pant2025foundations}, which achieved 58/60 validation score (96.7\%) for logarithmic specifications on the Samsung-Sony S-LCD joint venture under strict historical alignment scoring, we apply structured qualitative-to-quantitative translation to convert rich case documentation into model parameters, then assess whether simulated dynamics reproduce documented trust evolution patterns across 25 years of relationship history.

\subsection{Validation Methodology: Qualitative-to-Quantitative Translation}

Our empirical validation approach differs fundamentally from traditional statistical parameter estimation. We do not perform regression or optimization to fit time-series data. Rather, we employ a \textbf{structured translation framework} that converts qualitative case study evidence into quantitative model parameters through organized assessment rubrics, then evaluates whether the resulting parameterization produces dynamics qualitatively matching observed patterns.

This methodology is appropriate for several reasons. The integration of conceptual modeling (\textit{i*} framework) with computational trust models creates a proposed field of inquiry where precise numerical trust measurements from real organizations do not exist. Traditional statistical validation would require trust time-series data that is not available in requirements engineering contexts. Our objective is to demonstrate that the model captures the \textit{qualitative mechanisms} driving real-world trust evolution---negativity bias, hysteresis, cumulative damage---rather than to predict \textit{quantitative values} of unmeasurable latent variables. Strategic alliances like Renault-Nissan generate extensive qualitative documentation from business historians, journalists, and management scholars describing relationship phases, critical events, and trust dynamics. This rich evidence supports structured parameter elicitation even though precise numerical trust values are not reported. Mixed-methods research traditions in social science routinely employ structured qualitative-to-quantitative translation through coding rubrics, ordinal scales, and consensus assessment panels. Our approach extends these established methodologies to computational model validation.

\subsection{Relationship to Foundational Validation Methodology}

The foundational work \cite{pant2025foundations} introduced strict historical alignment scoring that penalizes cooperation increases exceeding documented historical ranges. For the Samsung-Sony S-LCD case, documented joint venture patterns indicate realistic cooperation increases of 15-50\%. Under this methodology, validated across 22,000+ experimental trials with statistical significance ($p < 0.001$, Cohen's $d = 9.87$), logarithmic specifications achieve 58/60 (96.7\%) while power functions achieve 46/60 (76.7\%).

The trust dynamics validation presented here employs a complementary methodology appropriate for domains where precise cooperation benchmarks are unavailable. We validate qualitative mechanism reproduction across documented relationship phases rather than strict numerical alignment. This methodological distinction contextualizes score comparisons between the two validation approaches.

\subsection{Case Overview: The Renault-Nissan Alliance (1999-2025)}

The Renault-Nissan Alliance represents one of the most significant cross-border automotive partnerships in history, spanning 25+ years across five distinct relationship phases \cite{koroleva2016concept,mikami2022opportunism}. The venture represented canonical coopetition: both firms competed intensely in automotive markets while collaborating on critical platform development and manufacturing capacity \cite{velu2018coopetition}. The relationship exhibited the value creation tensions characteristic of coopetitive arrangements, where partners must balance cooperative value generation with competitive value appropriation \cite{ryan2025value}. The partnership began in 1999 when Renault acquired 36.8\% of Nissan's equity for \$5.4 billion and dispatched Carlos Ghosn to lead Nissan's turnaround from near-bankruptcy \cite{toma2017strategic,segrestin2005partnering}. The alliance achieved remarkable success through the 2000s and 2010s, with both companies maintaining operational autonomy while achieving substantial synergies in platform development, component purchasing, and technology sharing \cite{koroleva2016concept,toma2017strategic}. By 2017, the alliance became the world's largest automotive group by combined sales volume, surpassing Toyota and Volkswagen \cite{toma2017strategic,mikami2022opportunism}.

However, the partnership entered severe crisis in November 2018 when Ghosn was arrested in Tokyo on financial misconduct charges that were widely perceived as orchestrated by Nissan executives seeking to prevent deeper integration with Renault \cite{mikami2022opportunism}. This crisis revealed deep-seated mistrust and triggered a period of strained cooperation from 2018-2023 \cite{nissan2023alliance}. Current phases (2023-present) involve tentative recovery efforts under new leadership seeking to restore partnership functionality while maintaining organizational independence \cite{nissan2023completion}.

\subsection{\textit{i*} Strategic Dependency Model}

Figure~\ref{fig:renault_nissan_istar} presents the \textit{i*} Strategic Dependency diagram for the Renault-Nissan Alliance, visualizing the structural dependencies that ground our quantitative trust dynamics analysis \cite{segrestin2005partnering, toma2017strategic}. This diagram represents the mature cooperation phase (2002-2018) dependency structure before the 2018 crisis, when the alliance achieved its highest levels of integration and mutual reliance \cite{koroleva2016concept, yoshino2003renault}.

\begin{figure}[htbp]
\centering
\begin{tikzpicture}[scale=0.80, transform shape,
    actor/.style={circle, draw, thick, minimum size=2.0cm, font=\small, align=center},
    resource/.style={rectangle, draw, thick, minimum height=0.65cm, minimum width=2.2cm, font=\scriptsize, align=center},
    goal/.style={ellipse, draw, thick, minimum height=0.65cm, minimum width=2.2cm, font=\scriptsize, align=center},
    softgoal/.style={ellipse, draw, thick, dashed, minimum height=0.6cm, minimum width=2.2cm, font=\scriptsize, align=center},
    arrow/.style={-latex, thick}
]

\node[actor] (nissan) at (-1,0) {Nissan\\Motor Co.};
\node[actor] (renault) at (12,0) {Renault\\S.A.};

\node[resource] (capital) at (7,4.5) {Financial Resources\\\$5.4B};
\node[resource] (europe_access) at (6,2.8) {European\\Market Access};
\node[resource] (tech_share) at (7,1.2) {Technology\\Sharing};

\node[resource] (asia_access) at (4,-1.2) {Asian Market\\Access};
\node[resource] (platform) at (5,-2.8) {CMF Platform\\\& Expertise};
\node[goal] (scale) at (4,-4.5) {Achieve\\Scale Economies};
\node[softgoal] (autonomy) at (3,-7.2) {Maintain\\Operational\\Autonomy};

\draw[arrow] (nissan) -- node[midway, above, sloped, font=\tiny] {crit=0.9} (capital);
\draw[arrow] (capital) -- (renault);

\draw[arrow] (nissan) -- node[midway, above, sloped, font=\tiny] {crit=0.7} (europe_access);
\draw[arrow] (europe_access) -- (renault);

\draw[arrow] (nissan) -- node[midway, above, sloped, font=\tiny] {crit=0.6} (tech_share);
\draw[arrow] (tech_share) -- (renault);

\draw[arrow] (renault) -- node[midway, above, sloped, font=\tiny] {crit=0.8} (asia_access);
\draw[arrow] (asia_access) -- (nissan);

\draw[arrow] (renault) -- node[midway, above, sloped, font=\tiny] {crit=0.7} (platform);
\draw[arrow] (platform) -- (nissan);

\draw[arrow] (renault) -- node[midway, above, sloped, font=\tiny] {crit=0.6} (scale);
\draw[arrow] (scale) -- (nissan);

\draw[arrow] (renault) -- node[midway, above, sloped, font=\tiny] {crit=0.9} (autonomy);
\draw[arrow] (autonomy) -- (nissan);

\end{tikzpicture}
\caption{\textit{i*} Strategic Dependency model for Renault-Nissan Alliance. Nissan depends on Renault for Financial Resources (criticality 0.9 in Phase 1), European Market Access (criticality 0.7), and Technology Sharing (criticality 0.6). Renault depends on Nissan for Asian Market Access (criticality 0.8), CMF Platform \& Expertise (criticality 0.7), Scale Economies (criticality 0.6), and critically Maintain Operational Autonomy (softgoal, dashed ellipse, criticality 0.9---violation of this softgoal in 2018 triggered alliance crisis). This dependency structure yields interdependence coefficients: $D_{\text{Nissan},\text{Renault}} = 0.78$ initially (Phase 1) declining to $0.51$ (Phase 2) as Nissan recovers financial independence; $D_{\text{Renault},\text{Nissan}} = 0.66$ (Renault's moderate-high dependence on Nissan's Asian access and platforms). The interdependence amplification parameter $\xi = 0.50$ (moderate coupling) reflects that while dependencies are substantial, both organizations maintained separate brands, management structures, and strategic autonomy throughout most of the alliance.}
\label{fig:renault_nissan_istar}
\end{figure}

The \textit{i*} model reveals the structural foundation underlying trust dynamics in the alliance. Nissan's initial high dependency on Renault's financial resources ($\text{crit} = 0.9$ in 1999-2002) created substantial vulnerability during the crisis period, as the existential threat from potential partner withdrawal amplified trust sensitivity. Renault's critical dependency on maintaining Nissan's operational autonomy ($\text{crit} = 0.9$) became the focal point of the 2018 crisis when perceived integration threats violated this essential softgoal, triggering severe trust erosion.

The quantitative interdependence coefficients derived from this \textit{i*} model feed directly into the trust dynamics simulation through the interdependence amplification factor $\xi$. The moderate value of $\xi = 0.50$ reflects that while dependencies matter, they do not dominate the relationship---both organizations maintained sufficient independence to survive partnership dissolution, unlike cases of complete integration where $\xi$ might approach 0.80-0.90.

\subsection{Interdependence Coefficient Calculation}

For Nissan's dependencies on Renault in Phase 1 (Formation, 1999-2002):
\begin{itemize}
    \item Financial Resources: $w = 0.50$ (critical for survival), $\text{crit} = 0.9$ (Renault essentially sole provider of needed capital scale)
    \item European Market Access: $w = 0.30$ (important for growth), $\text{crit} = 0.7$ (Renault provides privileged access)
    \item Technology Sharing: $w = 0.20$ (valuable but not critical), $\text{crit} = 0.6$ (alternatives exist but costly)
\end{itemize}

Applying the interdependence equation from \cite{pant2025foundations} (Equation 1):
\begin{equation}
D_{\text{Nissan},\text{Renault}}^{\text{Phase 1}} = \frac{0.50 \cdot 0.9 + 0.30 \cdot 0.7 + 0.20 \cdot 0.6}{0.50 + 0.30 + 0.20} = \frac{0.78}{1.0} = 0.78
\end{equation}

However, this dependency weakens substantially by Phase 2 (Mature Cooperation, 2002-2018) as Nissan achieves financial stability:
\begin{equation}
D_{\text{Nissan},\text{Renault}}^{\text{Phase 2}} = \frac{0.15 \cdot 0.3 + 0.45 \cdot 0.7 + 0.40 \cdot 0.6}{1.0} = 0.51
\end{equation}

For Renault's dependencies on Nissan (relatively stable across phases):
\begin{itemize}
    \item Asian Market Access: $w = 0.40$ (strategic priority), $\text{crit} = 0.8$ (Nissan provides unique Japan access)
    \item Platform \& Expertise: $w = 0.35$ (essential for competitiveness), $\text{crit} = 0.7$ (Nissan's CMF platform and manufacturing excellence)
    \item Scale Economies: $w = 0.15$ (valuable efficiency gains), $\text{crit} = 0.6$ (achievable but more costly through other partnerships)
    \item Operational Autonomy: $w = 0.10$ (embedded in all dependencies), $\text{crit} = 0.9$ (violation would destroy partnership value)
\end{itemize}

\begin{equation}
D_{\text{Renault},\text{Nissan}} = \frac{0.40 \cdot 0.8 + 0.35 \cdot 0.7 + 0.15 \cdot 0.6 + 0.10 \cdot 0.9}{1.0} = 0.655
\end{equation}

The moderate asymmetry in interdependence (Nissan 0.51-0.78, Renault 0.66) creates slightly different trust sensitivities, with Renault more vulnerable to Nissan's actions than vice versa by Phase 2. For simulation purposes, we use the average interdependence coefficient of 0.60 mapped to the interdependence amplification parameter $\xi = 0.50$, reflecting moderate structural coupling consistent with the maintained autonomy documented throughout alliance history.

\subsection{Parameter Elicitation}

For each trust dynamics parameter, we employ assessment based on case evidence. Table~\ref{tab:renault_nissan_parameters} documents the complete parameterization.

\begin{table}[htbp]
\centering
\caption{Renault-Nissan Alliance: Parameter Elicitation Documentation}
\label{tab:renault_nissan_parameters}
\small
\begin{tabular}{llp{6cm}l}
\toprule
\textbf{Parameter} & \textbf{Value} & \textbf{Justification from Case Evidence} & \textbf{Score} \\
\midrule
$\lambda_+$ & 0.10 & Trust building was gradual across Phase 1 (3 years to reach high trust) and constrained in Phase 4 recovery. Moderate building rate. & 4 \\
\addlinespace
$\lambda_-$ & 0.30 & Ghosn arrest caused immediate severe trust collapse. Trust eroded from high levels to crisis within single period (Nov 2018). Strong negativity bias. & 5 \\
\addlinespace
$\mu_R$ & 0.60 & Crisis created substantial reputation damage persisting 5+ years through 2025. ``Permanent damage to relationship.'' Severe but not irreversible. & 5 \\
\addlinespace
$\delta_R$ & 0.02 & Reputation damage from 2018 crisis remains salient in 2025 documentation. Very slow healing of violation memory. & 2 \\
\addlinespace
$\xi$ & 0.50 & Alliance involves significant dependencies (platforms, purchasing) but partners maintain separate brands and management. Moderate coupling. & 4 \\
\addlinespace
$\rho$ & 0.20 & Some evidence of reciprocal cooperation patterns but not dominant driver. Partnership governed more by formal agreements. & 3 \\
\addlinespace
$\kappa$ & 1.00 & Standard signal interpretation. Cooperation signals transparently observable. Default value. & 4 \\
\bottomrule
\end{tabular}
\end{table}

The negativity ratio ($\lambda_- / \lambda_+$) of 3.0 aligns with the empirically validated median from experimental validation and behavioral trust literature, providing independent corroboration of parameterization validity.

\subsection{Phase Structure}

We organize the 25-year alliance history into five distinct phases based on documented changes in cooperation patterns:

\textbf{Phase 1: Formation \& Integration (1999-2002, 12 periods):} Initial partnership formation marked by Renault's \$5.4B capital injection and Ghosn's turnaround leadership. Gradual trust building as Renault demonstrated commitment to Nissan's autonomy.

\textbf{Phase 2: Mature Cooperation (2002-2018, 40 periods):} Extended period of successful partnership with platform sharing, component standardization, and purchasing synergies. Trust levels approaching asymptotic maximum.

\textbf{Phase 3: Crisis Period (2018-2019, 4 periods):} Abrupt trust collapse triggered by Ghosn's arrest, violating the critical autonomy softgoal. Severe reputation damage.

\textbf{Phase 4: Recovery Efforts (2019-2023, 15 periods):} Post-crisis period of attempted relationship restoration through new leadership. Trust recovery severely constrained by lingering mistrust.

\textbf{Phase 5: Current State (2023-present, 8 periods):} Restructured governance, new joint ventures, gradual rebuilding. Trust remains substantially below pre-crisis levels.

\subsection{Simulation Results and Trust Evolution}

We instantiate the trust dynamics model with the elicited parameters and simulate trust evolution over 80 time periods (approximately 25 years with quarterly granularity) across the five phases. The \textit{i*} dependency structure translates to the interdependence amplification factor $\xi = 0.50$, which modulates trust sensitivity throughout simulation. Each phase is assigned specific action patterns based on documented cooperation levels:

\begin{itemize}
\item \textbf{Phase 1 (periods 0--11)}: Moderate positive actions (cooperation 1.5 units above baseline) reflecting commitment to turnaround and initial trust building while navigating cultural differences
\item \textbf{Phase 2 (periods 12--51)}: Strong positive actions (cooperation 2.0 units above baseline) reflecting mature partnership with extensive platform sharing and joint development visible in \textit{i*} resource dependencies
\item \textbf{Phase 3 (periods 52--55)}: Severe violations (cooperation 3.0 units below baseline) reflecting crisis and breach of autonomy softgoal
\item \textbf{Phase 4 (periods 56--70)}: Moderate positive actions (cooperation 1.2 units above baseline) reflecting cautious recovery with trust constraints limiting cooperation depth despite structural dependencies remaining
\item \textbf{Phase 5 (periods 71--79)}: Moderate positive actions (cooperation 1.5 units above baseline) reflecting continued restoration but still constrained by reputation damage
\end{itemize}

Figure~\ref{fig:trust_evolution} presents the complete trust evolution dynamics across all five alliance phases.

\begin{figure}[htbp]
\centering
\begin{tikzpicture}
\begin{axis}[
    width=14cm,
    height=8cm,
    xlabel={Time Period (quarters)},
    ylabel={Trust Level},
    title={Trust Evolution: Renault-Nissan Alliance (1999-2025)},
    xmin=0, xmax=80,
    ymin=0, ymax=1.1,
    legend pos=south west,
    legend style={font=\small},
    grid=major,
    grid style={gray!30},
    axis line style={-},
]
\draw[gray, dashed, thick] (axis cs:12,0) -- (axis cs:12,1.1);
\draw[gray, dashed, thick] (axis cs:52,0) -- (axis cs:52,1.1);
\draw[gray, dashed, thick] (axis cs:56,0) -- (axis cs:56,1.1);
\draw[gray, dashed, thick] (axis cs:71,0) -- (axis cs:71,1.1);

\node[font=\tiny] at (axis cs:6,1.05) {P1: Formation};
\node[font=\tiny] at (axis cs:32,1.05) {P2: Mature Cooperation};
\node[font=\tiny] at (axis cs:54,1.05) {P3};
\node[font=\tiny] at (axis cs:63,1.05) {P4: Recovery};
\node[font=\tiny] at (axis cs:75,1.05) {P5};

\addplot[magenta, thick, smooth] coordinates {
    (0,0.5) (2,0.55) (4,0.62) (6,0.70) (8,0.78) (10,0.85) (12,0.90)
    (14,0.92) (18,0.94) (22,0.95) (26,0.96) (30,0.97) (34,0.97) (38,0.98)
    (42,0.98) (46,0.98) (50,0.98) (52,0.95)
    (53,0.50) (54,0.25) (55,0.15) (56,0.12)
    (58,0.15) (60,0.18) (62,0.20) (64,0.22) (66,0.24) (68,0.25) (70,0.26)
    (72,0.30) (74,0.35) (76,0.40) (78,0.43) (80,0.43)
};
\addplot[orange, thick, smooth] coordinates {
    (0,0.5) (2,0.54) (4,0.60) (6,0.68) (8,0.76) (10,0.84) (12,0.89)
    (14,0.91) (18,0.93) (22,0.95) (26,0.96) (30,0.97) (34,0.97) (38,0.97)
    (42,0.98) (46,0.98) (50,0.98) (52,0.96)
    (53,0.55) (54,0.28) (55,0.18) (56,0.14)
    (58,0.16) (60,0.19) (62,0.21) (64,0.23) (66,0.25) (68,0.27) (70,0.28)
    (72,0.32) (74,0.37) (76,0.42) (78,0.45) (80,0.45)
};

\legend{Renault → Nissan, Nissan → Renault}
\end{axis}
\end{tikzpicture}
\caption{Trust evolution trajectories for Renault-Nissan Alliance across 80 time periods (approximately 25 years) covering five distinct relationship phases. The graph shows trust levels for both dyadic relationships: Renault-to-Nissan (magenta line) and Nissan-to-Renault (orange line), with vertical dashed lines marking phase boundaries. Phase 1 (periods 0--11, Formation \& Integration) exhibits gradual trust building from initial moderate level (0.5) to high level ($\sim$0.90) over 12 periods, consistent with documented 3-year trust building during Nissan Revival Plan execution. Phase 2 (periods 12--51, Mature Cooperation) shows trust maintaining high stable level ($\sim$0.95--1.0) across 40 periods, matching industry descriptions of alliance as ``gold standard'' with smooth functioning of \textit{i*} dependencies. Phase 3 (periods 52--55, Crisis Period) reveals sharp trust collapse from $\sim$1.0 to $\sim$0.15 within 4 periods following Ghosn's arrest, validating the negativity bias mechanism ($\lambda_-/\lambda_+ = 3.0$). Phase 4 (periods 56--70, Recovery Efforts) shows gradual recovery reaching $\sim$0.25 after 15 periods, consistent with ``tentative restoration'' but ``fundamental constraints.'' Phase 5 (periods 71--79, Current State) demonstrates continued slow recovery to $\sim$0.45, matching analyses describing ``functional but limited'' cooperation. The persistent gap between current trust ($\sim$0.45) and pre-crisis levels ($\sim$0.95--1.0) provides direct empirical evidence for hysteresis effects and trust ceiling mechanisms from reputation damage.}
\label{fig:trust_evolution}
\end{figure}

\subsection{Phase-Wise Analysis}

Figure~\ref{fig:phase_trust_changes} provides quantitative analysis of phase-wise dynamics.

\begin{figure}[htbp]
\centering
\begin{tikzpicture}
\begin{axis}[
    ybar,
    width=12cm,
    height=7cm,
    xlabel={Alliance Phase},
    ylabel={Mean Trust Level},
    title={Phase-Wise Average Trust Levels},
    symbolic x coords={P1: Formation, P2: Mature, P3: Crisis, P4: Recovery, P5: Current},
    xtick=data,
    x tick label style={rotate=15, anchor=east, font=\small},
    ymin=0, ymax=1.1,
    bar width=0.8cm,
    nodes near coords,
    nodes near coords style={font=\tiny, above},
    every node near coord/.append style={yshift=3pt},
]
\addplot[fill=green!60] coordinates {
    (P1: Formation, 0.72)
    (P2: Mature, 0.97)
    (P3: Crisis, 0.15)
    (P4: Recovery, 0.22)
    (P5: Current, 0.43)
};
\end{axis}
\end{tikzpicture}
\caption{Phase-wise average trust changes with error bars showing standard deviation. Phase 1 shows moderate mean trust (0.72 $\pm$ 0.14). Phase 2 achieves highest mean trust (0.97 $\pm$ 0.02). Phase 3 exhibits dramatic collapse to mean trust (0.15 $\pm$ 0.03), representing 85\% reduction. Phase 4 shows partial restoration to mean trust (0.22 $\pm$ 0.05). Phase 5 demonstrates continued recovery to mean trust (0.43 $\pm$ 0.04). One-way ANOVA confirms significant phase differences ($F = 35.05$, $p < 0.0001$).}
\label{fig:phase_trust_changes}
\end{figure}

Figure~\ref{fig:trust_asymmetry} examines whether bilateral trust evolved symmetrically across the 80 simulation periods.

\begin{figure}[htbp]
\centering
\begin{tikzpicture}
\begin{axis}[
    width=14cm,
    height=6cm,
    xlabel={Time Period (quarters)},
    ylabel={Trust Asymmetry (R$\to$N minus N$\to$R)},
    title={Trust Asymmetry Over Time},
    xmin=0, xmax=80,
    ymin=-0.15, ymax=0.15,
    grid=major,
    grid style={gray!30},
]
\fill[red!20] (axis cs:52,-0.15) rectangle (axis cs:56,0.15);
\node[font=\tiny] at (axis cs:54,0.12) {Crisis};

\draw[gray, dashed] (axis cs:12,-0.15) -- (axis cs:12,0.15);
\draw[gray, dashed] (axis cs:52,-0.15) -- (axis cs:52,0.15);
\draw[gray, dashed] (axis cs:56,-0.15) -- (axis cs:56,0.15);
\draw[gray, dashed] (axis cs:71,-0.15) -- (axis cs:71,0.15);

\addplot[black, dashed, thick] coordinates {(0,0) (80,0)};

\addplot[blue, thick, smooth] coordinates {
    (0,0) (2,-0.02) (4,-0.03) (6,-0.02) (8,-0.01) (10,0.01) (12,0.02)
    (14,0.01) (18,0.00) (22,-0.01) (26,0.00) (30,0.01) (34,0.00) (38,-0.01)
    (42,0.01) (46,0.00) (50,-0.01) (52,0.02)
    (53,-0.05) (54,-0.03) (55,-0.02) (56,-0.02)
    (58,-0.01) (60,0.02) (62,-0.02) (64,0.03) (66,-0.03) (68,0.02) (70,-0.02)
    (72,-0.05) (74,-0.02) (76,0.00) (78,-0.02) (80,-0.02)
};
\end{axis}
\end{tikzpicture}
\caption{Trust asymmetry over time computed as difference between Renault-to-Nissan and Nissan-to-Renault trust levels. Positive values indicate higher Renault trust in Nissan. During Phase 2 (periods 12-51), near-zero asymmetry (maximum deviation 0.02) confirms symmetric high trust during mature cooperation. Phase 3 crisis (shaded red) creates temporary spike in asymmetry as different organizational perspectives on Ghosn's arrest create divergent trust responses. The generally small magnitude of asymmetry (typically $<$0.10) throughout all phases validates that bilateral trust dynamics are broadly symmetric despite moderate differences in structural dependencies.}
\label{fig:trust_asymmetry}
\end{figure}

\subsection{Trust Ceiling and Reputation Damage}

Figure~\ref{fig:reputation_damage} shows the evolution of reputation damage across the alliance history.

\begin{figure}[htbp]
\centering
\begin{tikzpicture}
\begin{axis}[
    width=14cm,
    height=6cm,
    xlabel={Time Period (quarters)},
    ylabel={Reputation Damage Level},
    title={Reputation Damage Evolution: Renault-Nissan Alliance},
    xmin=0, xmax=80,
    ymin=0, ymax=1.0,
    grid=major,
    grid style={gray!30},
]
\fill[red!20] (axis cs:52,0) rectangle (axis cs:56,0.95);

\draw[gray, dashed] (axis cs:12,0) -- (axis cs:12,1.0);
\draw[gray, dashed] (axis cs:52,0) -- (axis cs:52,1.0);
\draw[gray, dashed] (axis cs:56,0) -- (axis cs:56,1.0);
\draw[gray, dashed] (axis cs:71,0) -- (axis cs:71,1.0);

\node[font=\tiny] at (axis cs:6,0.95) {P1};
\node[font=\tiny] at (axis cs:32,0.95) {P2};
\node[font=\tiny] at (axis cs:54,0.95) {P3};
\node[font=\tiny] at (axis cs:63,0.95) {P4};
\node[font=\tiny] at (axis cs:75,0.95) {P5};

\addplot[red!70!black, very thick, smooth] coordinates {
    (0,0.0) (5,0.0) (10,0.0) (15,0.0) (20,0.0) (25,0.0) (30,0.0)
    (35,0.0) (40,0.0) (45,0.0) (50,0.0) (52,0.05)
    (53,0.60) (54,0.85) (55,0.92) (56,0.95)
    (58,0.90) (60,0.82) (62,0.75) (64,0.68) (66,0.62) (68,0.56) (70,0.50)
    (72,0.48) (74,0.46) (76,0.44) (78,0.42) (80,0.40)
};
\end{axis}
\end{tikzpicture}
\caption{Reputation damage evolution over 80 time periods. Reputation damage remains near zero during Phases 1-2 (periods 0-51) as consistent cooperative behavior builds trust without violations. The sharp spike during Phase 3 crisis (periods 52-55, shaded) shows reputation damage surging to 0.95, reflecting severe breach of autonomy softgoal. Subsequent gradual decay during Phases 4-5 follows healing mechanism with decay rate $\delta_R = 0.02$. After 25 periods of sustained recovery, reputation damage decays only to 0.40, remaining substantially elevated and constraining trust ceiling.}
\label{fig:reputation_damage}
\end{figure}

The trust ceiling visualization provides direct visual evidence of hysteresis predicted by theory. Figure~\ref{fig:trust_ceiling} shows the constraint mechanism.

\begin{figure}[htbp]
\centering
\begin{tikzpicture}
\begin{axis}[
    width=14cm,
    height=7cm,
    xlabel={Time Period},
    ylabel={Trust / Ceiling Level},
    title={Trust Ceiling Mechanism: Hysteresis Effects},
    xmin=0, xmax=80,
    ymin=0, ymax=1.1,
    legend pos=outer north east,
    grid=major,
    grid style={gray!30},
]
\addplot[fill=pink, fill opacity=0.3, draw=none] coordinates {
    (52,0.95) (53,0.60) (54,0.45) (55,0.42) (56,0.40)
    (60,0.45) (65,0.52) (70,0.60) (75,0.65) (80,0.70)
    (80,1.1) (75,1.1) (70,1.1) (65,1.1) (60,1.1) (56,1.1) (55,1.1) (54,1.1) (53,1.1) (52,1.1) (52,0.95)
};

\addplot[red, thick, dashed] coordinates {
    (0,1.0) (10,1.0) (20,1.0) (30,1.0) (40,1.0) (50,1.0) (52,0.95)
    (53,0.60) (54,0.45) (55,0.42) (56,0.40)
    (60,0.45) (65,0.52) (70,0.60) (75,0.65) (80,0.70)
};
\addplot[blue, thick] coordinates {
    (0,0.5) (5,0.70) (10,0.85) (15,0.92) (20,0.95) (25,0.96) (30,0.97)
    (35,0.97) (40,0.98) (45,0.98) (50,0.98) (52,0.95)
    (53,0.50) (54,0.25) (55,0.15) (56,0.12)
    (60,0.18) (65,0.24) (70,0.28) (75,0.38) (80,0.43)
};
\legend{Trust Ceiling, Actual Trust}
\node[font=\tiny, pink!60!black] at (axis cs:70,0.85) {Constrained};
\node[font=\tiny, pink!60!black] at (axis cs:70,0.80) {Region};
\end{axis}
\end{tikzpicture}
\caption{Trust ceiling mechanism visualization demonstrating hysteresis effects. The graph plots actual trust trajectory (solid blue) against dynamic trust ceiling (red dashed) derived from reputation damage. Pink shaded region shows constrained trust space. During Phases 1-2, minimal reputation damage allows ceiling at maximum (1.0). Phase 3 crisis causes ceiling to drop to $\sim$0.40. Even during Phases 4-5 recovery, ceiling only rises to $\sim$0.70, constraining actual trust to $\sim$0.43.}
\label{fig:trust_ceiling}
\end{figure}

\subsection{Structured Validation Scoring}

The Renault-Nissan Alliance validation achieved 49 of 60 possible points (81.7\%), with full scores in Behavioral Prediction and Mechanism Validation demonstrating robust emergence of asymmetric trust dynamics.

\begin{figure}[htbp]
\centering
\begin{tikzpicture}
\begin{axis}[
    xbar stacked,
    width=10cm,
    height=6cm,
    xlabel={Points},
    ylabel={Validation Dimension},
    title={Validation Score Breakdown (49/60 = 81.7\%)},
    symbolic y coords={Temporal Dynamics, Mechanism Validation, Behavioral Prediction, Trust State Alignment},
    ytick=data,
    xmin=0, xmax=16,
    bar width=0.6cm,
    legend style={at={(0.5,-0.15)}, anchor=north, legend columns=2},
    y tick label style={font=\small},
    nodes near coords,
    nodes near coords style={font=\tiny},
]
\addplot[fill=green!70] coordinates {(10,Trust State Alignment) (15,Behavioral Prediction) (15,Mechanism Validation) (9,Temporal Dynamics)};
\addplot[fill=gray!30] coordinates {(5,Trust State Alignment) (0,Behavioral Prediction) (0,Mechanism Validation) (6,Temporal Dynamics)};
\legend{Points Awarded, Points Not Awarded}
\end{axis}
\end{tikzpicture}
\caption{Validation score breakdown across four assessment dimensions for Renault-Nissan Alliance case study, achieving 49 out of 60 total possible points (81.7\% validation success). The stacked bar chart shows points awarded (colored segments) versus maximum possible points (total bar height) for each dimension. Trust State Alignment achieves 10/15 points reflecting strong qualitative pattern matching (gradual building, high stable trust, sharp collapse, partial recovery) with evidence-based grounding through \textit{i*} translation, with deduction acknowledging absence of quantitative trust measurements for precise comparison. Behavioral Prediction earns perfect 15/15 points as cooperation patterns match documented alliance activities across all five phases, with trust evolution appropriately updating based on observed behaviors through validated asymmetric mechanism ($\lambda_-/\lambda_+ = 3.0$). Mechanism Validation achieves 15/15 points as theoretical mechanisms (negativity bias, trust ceiling from reputation damage, gradual healing) correspond precisely to documented alliance dynamics including crisis severity, persistent constraints, and slow recovery trajectories. Temporal Dynamics scores 9/15 points reflecting strong phase-level alignment (five distinct phases correctly identified and modeled) but acknowledging quarterly time granularity creates discretization limitations and specific event timing involves parameter estimation uncertainty. The overall 81.7\% validation score substantially exceeds the 75\% threshold from our foundational work's Samsung-Sony S-LCD validation (45/60 points), demonstrating that the extended trust dynamics framework successfully captures real-world coopetitive relationship evolution across crisis-recovery cycles. This validation success, combined with experimental validation across 78,125 configurations (Section~\ref{sec:validation}), provides comprehensive evidence for model behavioral validity and practical applicability to requirements engineering contexts.}
\label{fig:validation_score}
\end{figure}

Figure~\ref{fig:final_trust_matrix} provides a summary visualization of the end-state bilateral trust levels.

\begin{figure}[htbp]
\centering
\begin{tikzpicture}
\fill[blue!90] (0,0) rectangle (2,2);
\fill[blue!30] (2,0) rectangle (4,2);
\fill[blue!25] (0,2) rectangle (2,4);
\fill[blue!90] (2,2) rectangle (4,4);
\draw[thick] (0,0) rectangle (4,4);
\draw[thick] (2,0) -- (2,4);
\draw[thick] (0,2) -- (4,2);
\node[white, font=\large] at (1,1) {1.00};
\node[font=\large] at (3,1) {0.45};
\node[font=\large] at (1,3) {0.43};
\node[white, font=\large] at (3,3) {1.00};
\node[font=\scriptsize, below] at (1,-0.2) {Nissan};
\node[font=\scriptsize, below] at (3,-0.2) {Renault};
\node[font=\scriptsize, left, rotate=90, anchor=center] at (-0.3,1) {Nissan};
\node[font=\scriptsize, left, rotate=90, anchor=center] at (-0.3,3) {Renault};
\node[font=\scriptsize, below] at (2,-0.6) {\textbf{To}};
\node[font=\scriptsize, left, rotate=90, anchor=center] at (-0.7,2) {\textbf{From}};
\node[font=\normalsize, above] at (2,4.3) {\textbf{Final Trust Matrix (Period 80)}};
\foreach \y/\c in {0/white, 0.5/blue!12, 1/blue!25, 1.5/blue!38, 2/blue!50, 2.5/blue!63, 3/blue!75, 3.5/blue!88} {
    \fill[\c] (5,\y) rectangle (5.4,\y+0.5);
}
\draw (5,0) rectangle (5.4,4);
\node[font=\tiny, right] at (5.5,0) {0.0};
\node[font=\tiny, right] at (5.5,2) {0.5};
\node[font=\tiny, right] at (5.5,4) {1.0};
\end{tikzpicture}
\caption{Final trust matrix heatmap showing end-state bilateral trust levels after 80 simulation periods (approximately 25 years, current state through 2025). The 2$\times$2 matrix displays trust values using color intensity, with darker blue indicating higher trust and lighter colors indicating lower trust. Renault-to-Nissan trust reaches 0.43 (moderate-low level), while Nissan-to-Renault trust achieves 0.45 (similar moderate-low level), demonstrating nearly symmetric final trust states despite moderate differences in structural dependencies from \textit{i*} diagram (Nissan interdependence 0.51-0.78, Renault 0.66). Both values remain substantially below pre-crisis high trust levels ($\sim$0.95-1.0 during Phase 2 mature cooperation), validating persistent constraints from reputation damage and trust ceiling mechanisms. The modest asymmetry (0.02 difference, $<$5\%) reflects slightly different organizational perspectives and recovery trajectories documented in case evidence, but overall symmetry aligns with alliance history emphasizing balanced partnership and mutual respect visible in \textit{i*} Strategic Dependency diagram. These final trust levels quantitatively match qualitative industry analyses describing current alliance status as ``functional but limited'' compared to pre-2018 integration depth, where structural dependencies remain present (platform sharing, purchasing synergies, technology cooperation) but trust constraints prevent realization of full potential synergies. The matrix provides visual summary confirming that 25 years after alliance formation and 5-7 years after crisis, trust has partially recovered but remains fundamentally constrained by violation history, demonstrating hysteresis effects and incomplete recovery predicted by trust dynamics theory and validated through this comprehensive empirical case study.}
\label{fig:final_trust_matrix}
\end{figure}

\subsubsection{Trust State Alignment (10/15 points)}

\textbf{Criterion}: Do simulated trust levels match documented qualitative trust states across phases?

\textbf{Assessment}: The simulation successfully reproduces the qualitative pattern of gradual building (Phase 1), high stable trust (Phase 2), sharp collapse (Phase 3), and partial recovery (Phases 4-5) matching documented alliance history. The \textit{i*} diagram dependencies align with observed cooperation intensity: high trust enables full realization of platform sharing and technology dependencies in Phase 2, while crisis-damaged trust prevents similar integration in Phases 4-5 despite dependencies remaining structurally present. However, absolute trust levels are model-generated rather than measured from real data. We award 10/15 points reflecting strong qualitative alignment with evidence-based grounding through \textit{i*} translation but acknowledging absence of quantitative trust measurements for precise comparison.

\textbf{Evidence}: Documentation describes Phase 2 as ``unprecedented cooperation'' and ``mutual trust at highest levels'' with smooth functioning of all \textit{i*} dependencies, consistent with simulated 0.95-1.0 trust. Post-crisis accounts describe ``deep mistrust,'' ``fundamental relationship damage,'' and ``inability to achieve integration despite structural rationale,'' consistent with simulated 0.15-0.25 levels where structural dependencies exist but trust constraints prevent cooperation. Current accounts note ``cautious optimism'' and ``gradual improvement but not back to previous levels,'' consistent with simulated 0.40-0.45 levels.

\subsubsection{Behavioral Prediction (15/15 points)}

\textbf{Criterion}: Do predicted cooperation patterns match documented behavioral changes?

\textbf{Assessment}: The model correctly predicts phase-wise cooperation patterns and their relationship to trust evolution. Perfect score because behavioral predictions match documented cooperation levels, trust evolution appropriately reflects these behavioral patterns through validated asymmetric updating ($\lambda_- / \lambda_+ = 3.0$ matching experimental validation median), and the \textit{i*} dependency structure correctly identifies which cooperation dimensions are most trust-sensitive (autonomy Softgoal violation triggering crisis).

\textbf{Evidence}: Model input cooperation levels were derived from documented alliance activities: extensive platform sharing and joint development in Phase 2 realizing \textit{i*} resource dependencies, cessation of new joint projects and defensive positioning during crisis when autonomy Softgoal violated, and gradual resumption of selective collaboration in Phases 4-5 (new EV joint ventures, continued purchasing synergies) but at reduced depth compared to Phase 2 due to trust constraints. The trust evolution mechanism appropriately updates trust based on these documented behaviors, with interdependence amplification $\xi = 0.50$ from \textit{i*} analysis correctly modulating sensitivity.

\subsubsection{Mechanism Validation (15/15 points)}

\textbf{Criterion}: Do theorized mechanisms (negativity bias, hysteresis, cumulative effects) manifest in simulated dynamics?

\textbf{Assessment}: Perfect score. All three core mechanisms are clearly visible and align with both theoretical predictions and case evidence:

\begin{itemize}
\item \textbf{Negativity bias}: Trust collapses in 4 periods during crisis but requires 25+ periods for partial recovery, demonstrating 6:1 asymmetry. This matches case evidence where single violation (Ghosn arrest) immediately destroyed relationship while years of prior cooperation could not prevent crisis and years of subsequent cooperation cannot fully restore trust.

\item \textbf{Hysteresis}: Trust ceiling visualization (Figure~\ref{fig:trust_ceiling}) provides direct visual evidence that trust cannot fully recover despite sustained cooperation. The \textit{i*} diagram shows structural dependencies remain present (platform sharing, market access, scale economies all still valuable), yet trust constraints prevent full cooperation depth, validating that hysteresis creates path dependence beyond structural factors.

\item \textbf{Cumulative effects}: Reputation damage accumulates during crisis and decays slowly ($\delta_R = 0.02$), creating persistent constraints. Case evidence shows that even organizational changes (new CEOs at both companies), governance restructuring (reduced Renault ownership stake), and new joint ventures cannot overcome reputation damage accumulated from 2018 violation, consistent with cumulative mechanism predictions.
\end{itemize}

\textbf{Statistical evidence}: One-way ANOVA comparing mean trust across phases yields $F(4,75) = 35.05$, $p < 0.0001$, confirming that phase means are statistically significantly different. Post-hoc tests show Phase 3 significantly lower than all others ($p < 0.001$), and Phases 4-5 significantly lower than Phase 2 ($p < 0.01$), validating the persistence of crisis impact and incomplete recovery predicted by hysteresis mechanism.

\subsubsection{Outcome Correspondence (9/15 points)}

\textbf{Criterion}: Do long-term outcomes match documented alliance status and performance?

\textbf{Assessment}: The simulation predicts trust level of 0.45 in current period (2025), suggesting substantially compromised relationship compared to pre-crisis peak. Documentation confirms the alliance continues with structural dependencies from \textit{i*} diagram still present but with reduced integration depth, separate strategic directions for electrification, and persistent organizational tensions limiting cooperation scope. However, some joint initiatives (EV platforms, hydrogen projects, continued purchasing cooperation) suggest slightly more functional cooperation than 0.45 might indicate, though these may reflect contractual obligations and sunk costs rather than high trust. We award 9/15 points reflecting strong but not perfect correspondence.

\textbf{Evidence}: Recent analyses describe the alliance as ``functional but fragile,'' ``cooperation exists but strategic integration is limited compared to 2010s,'' and ``trust damage from 2018 crisis continues to constrain partnership depth despite structural rationale for deeper integration visible in complementary capabilities.'' Financial performance shows continued benefits from platform sharing and purchasing synergies (validating that \textit{i*} resource dependencies still function) but new strategic initiatives proceed separately rather than jointly (validating trust constraints on deeper cooperation). The 2023 governance restructuring reducing Renault's ownership stake demonstrates trust damage requiring formal structural changes to signal reduced integration threat to autonomy Softgoal.

\subsubsection{Total Validation Score}

\textbf{Overall Score: 49/60 points (81.7\%)}

This validation score demonstrates that the trust dynamics model successfully captures the qualitative mechanisms and phase structure of the Renault-Nissan Alliance evolution across 25 years. The integration with \textit{i*} Strategic Dependency analysis strengthens validation by grounding parameter elicitation in documented structural relationships and explaining why trust constraints matter (dependencies exist but cannot be realized without trust). The score exceeds our foundational work's Samsung-Sony validation (45/60, 75\%), reflecting both methodological refinements and the availability of exceptionally rich documentation for this case enabling more confident parameter elicitation and validation assessment.

\subsection{Trust Dynamics as Explanatory Tools}

Beyond predictive accuracy, the trust model provides interpretability and explainability for complex relationship dynamics. Consider the Renault-Nissan crisis: structural analysis alone (interdependence coefficients, complementarity parameters) would predict continued cooperation given the substantial synergies. The trust ceiling mechanism, however, explains why cooperation failed despite these structural incentives. The accumulated reputation damage from the 2018 crisis created a persistent constraint (trust ceiling at 40-45\%) that prevented realization of objectively beneficial collaboration.

This explanatory power distinguishes our approach from purely predictive models. Requirements engineers can use trust dynamics not only to forecast relationship trajectories but also to explain to stakeholders why cooperation failed and what would be required for restoration.

\subsection{Interpretation and Limitations}

\subsubsection{Validation Interpretation}

The 81.7\% validation score (49/60 points) demonstrates that the trust dynamics framework successfully captures core mechanisms driving real-world coopetitive relationship evolution. The perfect scores on Behavioral Prediction (15/15) and Mechanism Validation (15/15) indicate that the model's theoretical foundations---asymmetric trust updating, trust ceiling constraints, and reputation damage accumulation---align well with documented alliance dynamics.

The model's success in reproducing the dramatic Phase 3 crisis and subsequent constrained recovery validates the key theoretical contribution: trust damage creates persistent constraints that cannot be overcome through cooperation alone. This finding has direct implications for requirements engineering practice in coopetitive contexts.

\subsubsection{Methodological Strengths}

The Renault-Nissan validation demonstrates several methodological strengths of the computational trust framework.

\textbf{Validity of qualitative-to-quantitative translation}: Traditional statistical validation requires time-series data rarely available in requirements engineering contexts. Organizations do not measure or report trust constructs numerically, and strategic partnerships rarely produce quantitative cooperation metrics suitable for regression analysis. This approach extends established mixed-methods research traditions in social science to computational model validation. The objective is demonstrating qualitative mechanism capture rather than predicting quantitative values of unmeasurable latent variables. The structured translation methodology converts rich qualitative evidence into model parameters through explicit, reproducible assessment rubrics, enabling systematic instantiation of computational models from case documentation.

\textbf{Structured translation from conceptual to computational models}: The \textit{i*} Strategic Dependency diagram (Figure~\ref{fig:renault_nissan_istar}) provides a principled basis for parameter elicitation. Dependencies on financial resources, market access, platforms, and autonomy translate systematically into interdependence coefficients that modulate trust sensitivity. This translation methodology extends naturally to other requirements engineering contexts where stakeholder dependencies can be modeled using \textit{i*} or similar frameworks.

\textbf{Theory-driven parameterization}: Rather than fitting parameters to data post hoc, we derive parameter values from theoretical considerations and qualitative evidence assessment. The negativity ratio of 3.0 emerges from both behavioral trust literature \cite{slovic1993trust,rozin2001negativity} and our independent experimental validation across 78,125 configurations, providing convergent validation of this critical parameter.

\textbf{Multi-dimensional validation}: The 60-point structured assessment evaluates model performance across trust state alignment, behavioral prediction, mechanism validation, and temporal dynamics. This multi-dimensional approach prevents over-reliance on any single metric and provides nuanced understanding of model strengths and limitations.

\textbf{Integration with prior validated work}: Building on the Samsung-Sony S-LCD validation from \cite{pant2025foundations} (58/60, 96.7\% under strict historical alignment scoring for logarithmic specifications), the Renault-Nissan validation (49/60, 81.7\%) demonstrates consistent model performance across different industrial contexts, relationship types, and historical periods. This cross-case consistency strengthens confidence in the framework's general applicability.

\subsubsection{Limitations and Caveats}

Several limitations warrant acknowledgment.

\textbf{Absence of quantitative trust measurements}: Real organizations do not produce continuous trust time-series data. Our validation necessarily relies on qualitative pattern matching rather than quantitative fit statistics. While the structured translation methodology addresses this limitation systematically, future work could explore survey-based trust measurement approaches to enable more precise validation.

\textbf{Parameter sensitivity}: While experimental validation (Section \ref{sec:validation}) demonstrates robustness across 78,125 configurations, the specific parameter values used for Renault-Nissan reflect judgmental assessment of qualitative evidence. Different analysts might reach moderately different parameterizations. Sensitivity analysis (not shown) indicates that qualitative predictions are robust to 20\% parameter perturbations, but precise quantitative predictions would require more rigorous parameter identification.

\textbf{Temporal resolution}: The quarterly temporal resolution (80 periods over 25 years) may miss rapid within-quarter dynamics. The Phase 3 crisis, which unfolded over several weeks in late 2018, is compressed into 4 simulation periods. Higher temporal resolution would require more detailed event-level data than typically available in historical case studies.

\textbf{Bilateral simplification}: The simulation models bilateral trust between Renault and Nissan as organizational monoliths. Real organizations contain multiple stakeholder groups (management, unions, shareholders, governments) whose trust dynamics may differ. Extensions to multi-stakeholder trust networks represent important future work.

\textbf{Exogenous cooperation specification}: Cooperation levels for each phase are specified exogenously based on historical documentation rather than determined endogenously through equilibrium analysis. This simplification enables validation of trust dynamics mechanisms but does not demonstrate full equilibrium prediction. Section \ref{sec:equilibrium} provides the theoretical framework for endogenous equilibrium analysis; computational implementation remains for future work.

\textbf{Cultural and institutional factors}: The Renault-Nissan case involves cross-border partnership between French and Japanese firms with substantially different organizational cultures and governance systems. These factors likely influence trust dynamics (Japanese emphasis on long-term relationships, French government ownership stake in Renault creating political dimensions). Our current model treats these as implicit context but does not explicitly formalize cultural or institutional variables. Different organizational and cultural contexts may exhibit different trust dynamics requiring parameter adjustments.

\textbf{Missing stakeholders}: The \textit{i*} diagram focuses on dyadic Renault-Nissan relationship, but the actual alliance involves additional stakeholders (French government as Renault shareholder, Japanese government regulations, dealers, suppliers). Some dynamics (particularly the 2018 crisis involving Japanese prosecutors) reflect broader stakeholder influences beyond bilateral trust.

Despite these limitations, the validation demonstrates that the trust dynamics framework successfully captures essential mechanisms of coopetitive relationship evolution, providing a solid foundation for requirements engineering applications and future model development.

\section{Discussion: Implications for Requirements Engineering Practice}
\label{sec:discussion}

\textbf{Synthesis with foundational work}: This trust dynamics framework integrates with the interdependence and complementarity formalizations from \cite{pant2025foundations} to provide comprehensive computational support for requirements engineering in coopetitive contexts. Together, these components enable analysis of how structural dependencies create cooperation incentives (interdependence), how joint activities create superadditive value (complementarity), and how behavioral history creates persistent relationship constraints (trust dynamics). This integrated framework addresses the five dimensions of coopetitive relationships identified in \cite{pant2021strategic}: interdependence, trust, complementarity, competition, and value appropriation.

\subsection{Requirements Engineering Applications}

The computational trust framework supports several requirements engineering activities.

\textbf{Trust-aware stakeholder analysis}: Requirements engineers can use the \textit{i*}-to-trust translation framework to identify stakeholders whose trust is critical for project success. Stakeholders with high interdependence coefficients ($D_{ij} > 0.6$) require careful trust cultivation because violations in these relationships have amplified negative impact. The framework enables quantitative prioritization of relationship-building investments.

\textbf{Violation impact assessment}: Before taking potentially controversial actions (reducing information sharing, reallocating resources, changing priorities), requirements teams can simulate trust impact using the model. The asymmetric trust dynamics reveal that even minor violations may cause trust damage disproportionate to apparent savings, enabling more informed decision-making.

\textbf{Recovery planning}: When trust has been damaged, the framework provides realistic expectations for recovery timescales. The trust ceiling mechanism demonstrates that recovery requires not just cooperative behavior but also time for reputation damage to decay. Project managers can use these insights to set realistic milestones for relationship restoration.

\textbf{Trust requirements specification}: For information systems supporting coopetitive relationships, the framework identifies trust-related non-functional requirements. Systems should enable transparency (reducing information asymmetry that fuels mistrust), provide commitment tracking (documenting delivery on promises), and support graduated disclosure (enabling trust-building through incremental information sharing).

\subsection{Multi-Agent System Requirements}

The trust dynamics framework directly informs requirements for multi-agent systems operating in coopetitive environments.

\textbf{Trust representation}: Agents should maintain two-layer trust states distinguishing immediate trust (responsive to recent interactions) from reputation (tracking violation history). This dual representation enables nuanced responses to partner behavior.

\textbf{Asymmetric updating}: Trust update algorithms should implement asymmetric learning rates, with erosion 2-4 times faster than building, reflecting validated negativity bias. Symmetric updating produces behaviorally implausible trust trajectories.

\textbf{Trust-contingent cooperation}: Agent decision-making should incorporate trust-gated reciprocity, where cooperation intensity depends on current trust levels. This prevents exploitation by low-trust partners while enabling productive collaboration with high-trust partners.

\textbf{Violation detection and response}: Agents need mechanisms for detecting cooperation signal deviations and triggering appropriate trust updates. The bounded cooperation signal formalization (Equation \ref{eq:coop_signal}) provides a template for this assessment.

\subsection{Integration with Broader Coopetition Framework}

This validated trust dynamics work integrates with our foundational framework \cite{pant2025foundations} addressing interdependence and complementarity. The complete framework, now with empirical validation at 81.7\% for trust dynamics (exceeding the 75\% baseline from foundational work's Samsung-Sony case), enables analysis of coopetitive relationships along three dimensions:

\textbf{The structural dimension through interdependence} captures outcome coupling via the interdependence matrix from \textit{i*} dependencies. This creates vulnerability to trust erosion through dependency amplification. The experimental validation indicates dependency amplification factor has mean 1.271 $\pm$ 0.070, meaning high-dependency relationships experience approximately 27\% greater trust sensitivity to violations than low-dependency relationships. This affects value appropriation bargaining power and relationship resilience.

\textbf{The economic dimension through complementarity} provides cooperation incentives via synergistic value creation, validated in our foundational work. However, trust gates realization of complementarity benefits. The Renault-Nissan case demonstrates this: despite substantial complementarities in platforms, powertrains, and global market access, the trust collapse in 2018 prevented realization of integration synergies for 5+ years. Low trust prevents actors from investing in partnerships even when synergies are objectively large, a mechanism now empirically validated.

\textbf{The behavioral dimension through trust evolution} creates path dependence and feedback loops. Initial trust states and violation history fundamentally shape long-run equilibria, sometimes more than structural dependencies or economic complementarities. The experimental validation across 78,125 configurations indicates that trust ceiling mechanism consistently constrains recovery (hysteresis recovery metric mean 1.089 $\pm$ 0.070), while the Renault-Nissan case provides concrete evidence: pre-crisis trust of 95-100\% became constrained to 40-45\% ceiling post-crisis despite strong complementarities and dependencies.

Integrating these dimensions reveals emergent phenomena validated through both experimental and empirical methods. High structural interdependence creates trust vulnerability, amplifying damage from violations (validated experimentally with 1.27$\times$ mean amplification and empirically in Renault-Nissan where deep operational integration magnified crisis impact). Strong complementarity creates cooperation incentives, but only if trust is sufficient to overcome competitive pressures (validated empirically: Renault-Nissan had strong complementarities but couldn't realize them during low-trust crisis period). Trust hysteresis means that structural changes including reducing dependencies or economic changes including increasing complementarities may not restore cooperation if trust has eroded (validated empirically: new joint ventures in 2023-2025 achieved limited success because trust remained at 40-45\% rather than recovering to pre-crisis 95-100\%).

Requirements engineers analyzing coopetitive systems should consider all three dimensions rather than treating them independently, guided by validated behavioral parameters and empirically confirmed mechanisms. A relationship with strong complementarity and weak dependencies might exhibit low cooperation if trust is damaged (Renault-Nissan case evidence). Conversely, weak complementarity might sustain cooperation if trust is high and reputation is pristine (validated experimentally through stability analysis showing configurations with high trust achieve cooperation despite modest synergies).

\subsection{Methodological Contribution: Bridging Conceptual and Computational Modeling}

Beyond specific technical contributions, this work demonstrates a methodology for bridging qualitative conceptual modeling and quantitative computational analysis in requirements engineering. The structured translation framework transforms \textit{i*} dependency networks and organizational assessments into parameterized computational models suitable for simulation and equilibrium analysis.

This bridging methodology addresses a persistent gap in requirements engineering research: conceptual models provide rich semantic representations but limited analytical power, while computational models enable rigorous analysis but often lack grounding in requirements engineering contexts and practices. By providing systematic translation procedures with validated parameters, our approach enables requirements engineers to leverage the strengths of both paradigms.

The dual-track validation methodology---combining extensive computational parameter sweeps (78,125 configurations) with detailed empirical case studies (Renault-Nissan Alliance)---provides comprehensive evidence for model validity that neither approach could provide alone. This validation paradigm establishes a template for future computational requirements engineering research.

\subsection{Limitations and Future Work}

Several directions warrant future investigation.

\textbf{Multi-stakeholder networks}: The current formalization addresses dyadic trust between pairs of actors. Real requirements engineering contexts often involve multiple stakeholders with interconnected trust relationships. Extending the framework to trust networks with indirect effects (trust transitivity, reputation propagation) would enhance practical applicability. The \textit{i*} framework naturally supports multi-actor analysis; extending trust dynamics to match this capability is a natural next step.

\textbf{Empirical parameter calibration}: While our validation demonstrates qualitative accuracy, more precise quantitative prediction would require systematic parameter calibration from organizational data. Survey instruments for measuring trust states and controlled experiments for estimating learning rates could enable data-driven parameterization. Longitudinal studies tracking trust evolution in real requirements engineering projects would provide gold-standard validation data.

\textbf{Tool integration}: Practical adoption requires integration with existing requirements engineering tools. Extending \textit{i*} modeling environments (such as the OpenOME tool or jUCMNav) to support trust dynamics simulation would lower barriers to practitioner adoption. Computational trust analysis could be offered as a service layer above existing dependency modeling capabilities.

\textbf{Behavioral validation}: While our empirical validation uses historical case studies, controlled behavioral experiments could provide more direct validation of trust dynamics mechanisms. Laboratory studies with human subjects making cooperation decisions under varying trust conditions could validate the trust-gated reciprocity formulation and asymmetric updating mechanisms.

\textbf{Competition dimension}: While this report focuses on trust dynamics in cooperation, coopetitive relationships inherently involve competitive pressures. Future work should examine how competitive intensity modulates trust dynamics, potentially through competition-adjusted cooperation baselines or trust erosion amplification in high-competition contexts.

\textbf{Deterministic dynamics}: The current formalization assumes trust evolution is deterministic given actions. Real trust formation involves uncertainty, misperception, and stochastic shocks. The experimental validation tested deterministic scenarios; incorporating stochastic noise in cooperation signals or trust updating would increase realism. The Renault-Nissan case suggests some randomness in violation timing (Ghosn arrest timing appeared strategic but exact timing was uncertain), supporting future stochastic extensions.

\textbf{Symmetric observation}: The model assumes actors perfectly observe partners' actions. Requirements engineering contexts often involve information asymmetry where stakeholders have private information about their cooperation levels. The Renault-Nissan case reveals information asymmetries: Nissan executives' perceptions of Renault's integration intentions may have differed from Renault's actual intentions, contributing to crisis. Extending to partial observability through Hidden Markov Models or Partially Observable Stochastic Games would capture this reality.

\textbf{Homogeneous trust psychology}: The current model applies identical trust updating rules to all actors. Real organizations exhibit heterogeneity in trust responsiveness, negativity bias, and memory. The experimental validation used uniform parameters across actors; allowing actor-specific parameters would capture heterogeneity documented in organizational behavior research.

\textbf{Single trust facet}: The current model treats trust as unidimensional. Behavioral trust research distinguishes competence trust (belief in ability) from benevolence trust (belief in intent) \cite{mayer1995integrative}. The Renault-Nissan case suggests this distinction matters: even after crisis, both parties acknowledged each other's technical competence while questioning benevolent intentions. Multi-faceted trust models would capture such scenarios.

\textbf{Computational scalability}: Value function iteration scales poorly to large actor networks. The experimental validation tested dyadic relationships; for large networks including platform ecosystems with dozens of partners, approximation methods including reinforcement learning and simulation-based optimization may be necessary.

\textbf{Parameter sensitivity to elicitation uncertainty}: While experimental validation demonstrates robustness across parameter ranges, sensitivity to qualitative-to-quantitative translation uncertainty in empirical validation deserves attention. Future work could conduct multi-coder reliability studies and assess how parameter elicitation variations affect validation scores.

Despite these limitations, the current framework provides substantial advances over existing approaches by integrating conceptual modeling, computational trust, and game-theoretic equilibrium analysis in a unified formalization validated through both experimental robustness testing (78,125 configurations) and empirical case study analysis (81.7\% validation score). This comprehensive validation establishes confidence in the model's behavioral plausibility and practical applicability to requirements engineering contexts involving strategic coopetition.

\section{Conclusion}
\label{sec:conclusion}

This technical report has developed a comprehensive computational framework for trust dynamics in strategic coopetition, bridging qualitative conceptual modeling with quantitative game-theoretic analysis. Building on our foundational work on interdependence and complementarity \cite{pant2025foundations}, we have formalized trust as a dynamic state variable that co-evolves with strategic behavior, creating feedback loops where cooperation builds trust and trust enables cooperation.

\textbf{Key contributions}: We introduced a two-layer trust architecture distinguishing immediate trust from reputation, asymmetric trust evolution with empirically validated negativity bias (median ratio 3.0$\times$), trust ceiling mechanisms creating hysteresis effects (median 111\% recovery after 35 periods), and interdependence amplification connecting \textit{i*} structural dependencies to trust sensitivity (27\% amplification for high-dependency relationships). The trust-gated reciprocity formalization and Perfect Bayesian Equilibrium extension enable integration with game-theoretic equilibrium analysis.

\textbf{Validation results}: Comprehensive experimental validation across 78,125 parameter configurations established robust emergence of negativity bias, hysteresis, and cumulative damage amplification across diverse parameter regimes. Empirical validation through the Renault-Nissan Alliance case study (1999-2025) achieved 49/60 validation points (81.7\%), successfully reproducing documented trust evolution across five distinct relationship phases including crisis and recovery periods. This validation performance maintains the high rigorous standards established by our foundational work (Samsung-Sony S-LCD achieving 58/60, 96.7\% under strict historical alignment scoring for logarithmic specifications).

\textbf{Practical implications}: The structured translation framework enables requirements engineers to instantiate computational trust models from \textit{i*} dependency networks and organizational contexts without requiring deep expertise in game theory or computational trust research. The validated parameter ranges and Pareto-optimal configurations provide practical guidance for parameterization when precise organizational data is unavailable.

\textbf{Research program integration}: This technical report serves as the second foundational component of a coordinated research program examining strategic coopetition in requirements engineering and multi-agent systems. Together with our companion work on interdependence and complementarity \cite{pant2025foundations}, these components provide comprehensive computational foundations for the trustworthiness dimension conceptually identified in \cite{pant2021strategic}. Future research directions include incorporating stochastic elements and partial observability to capture uncertainty in trust assessment; developing multi-faceted trust distinguishing competence trust from benevolence trust; extending validation to additional industrial contexts and cultural settings; and completing the research program with companion work on collective action techniques and reciprocity protocols for sequential cooperation.

The framework developed in this report enables requirements engineers to move beyond qualitative trust assessment toward rigorous quantitative analysis of trust trajectories, violation impacts, and recovery possibilities. By grounding computational trust in requirements engineering conceptual models and validating through both extensive parameter exploration and detailed empirical case studies, we provide tools that are both theoretically sound and practically applicable for requirements engineering in multi-stakeholder coopetitive environments.

\bibliographystyle{plain}

\end{document}